\documentclass[10pt,journal,compsoc]{IEEEtran}
\usepackage{lineno,hyperref}
\usepackage { amsmath , amssymb , amsthm }
\usepackage [ utf8 ]{ inputenc }
\usepackage{caption, subcaption}
\usepackage{tabularx}
\usepackage{multirow}
\usepackage[linesnumbered,ruled,vlined]{algorithm2e}
\usepackage{xcolor}
\usepackage{verbatim}
\usepackage{enumitem}
\usepackage[normalem]{ulem} 

\ifCLASSOPTIONcompsoc
  \usepackage[nocompress]{cite}
\else
  \usepackage{cite}
\fi
\ifCLASSINFOpdf
   \usepackage[pdftex]{graphicx}
\else
\fi
\begin{document}
%
\title{Boosting Graph Embedding on a Single GPU}
%
%
%

\author{Amro~Alabsi~Aljundi,
        Taha~Atahan~Akyildiz,
        and~Kamer~Kaya
\IEEEcompsocitemizethanks{
\IEEEcompsocthanksitem A. Alabsi Aljundi, A. Akyildiz, and K. Kaya are with the Faculty of Engineering and Natural Sciences, Sabancı University, Istanbul, Turkey.\\ 
Email: \{amroa, aakyildiz, kaya\}@sabanciuniv.edu}%
\thanks{The first two authors contributed equally to the paper.}%
\thanks{A preliminary version is appeared in ``ICPP '20: 49th Int. Conf. on Par. Processing" as ``GOSH: Embedding Big Graphs on Small Hardware".}}%

%
%

\markboth{IEEE TRANSACTIONS ON PARALLEL AND DISTRIBUTED COMPUTING}%
{Boosting Graph Embedding on a Single GPU} 
%



\newcommand{\smalltrain}{{\sc TrainInGPU}\xspace}
\newcommand{\coa}{{\sc MultiEdgeCollapse}\xspace}
\newcommand{\lge}{{\sc LargeGraphGPU}\xspace}
\newcommand{\malgo}{{\sc Gosh}\xspace}
\newcommand{\malgow}{{\sc GoshW}\xspace}
\newcommand{\versex}{{\sc {Verse}}\xspace}
\newcommand{\linex}{{\sc {Line}}\xspace}
\newcommand{\graphvite}{{\sc {GV}}\xspace}
\newcommand{\graphvitee}{{\sc {Graphvite}}\xspace}
\newcommand{\mile}{{\sc {Mile}}\xspace}
\newcommand{\pbg}{{\sc {PBG}}\xspace}
\newcommand{\pbggg}{{\sc {PyTorch-BigGraph}}\xspace}
\newcommand{\pbgg}{{\sc {PyTorch-BigG}}\xspace}
\newcommand{\harp}{{\sc {Harp}}\xspace}

\newcommand\new[1]{\textcolor{blue}{#1}}
\newcommand\added[1]{\textcolor{purple}{#1}}
\newcommand\unclear[1]{\textcolor{red}{#1}}
\SetKwRepeat{Do}{do}{while}

\IEEEtitleabstractindextext{%
\begin{abstract}
Graphs are ubiquitous, and they can model unique characteristics and complex relations of real-life systems. Although using machine learning (ML) on graphs is promising, their raw representation is not suitable for ML algorithms. Graph embedding represents each node of a graph as a $d$-dimensional vector which is more suitable for ML tasks. 
However, the embedding process is expensive, and CPU-based tools do not scale to real-world graphs. In this work, we present GOSH, a GPU-based tool for embedding large-scale graphs with minimum hardware constraints. GOSH employs a novel graph coarsening algorithm to enhance the impact of updates and minimize the work for embedding. It also incorporates a decomposition schema that enables any arbitrarily large graph to be embedded with a single GPU. As a result, GOSH sets a new state-of-the-art in link prediction both in accuracy and speed, and delivers high-quality embeddings for node classification at a fraction of the time compared to the state-of-the-art. For instance, it can embed a graph with over 65 million vertices and 1.8 billion edges in less than 30 minutes on a single GPU.
\end{abstract}

}

\maketitle
\IEEEdisplaynontitleabstractindextext

%
\IEEEpeerreviewmaketitle

\ifCLASSOPTIONcompsoc
\IEEEraisesectionheading{\section{Introduction}\label{sec:introduction}}
\else
\section{Introduction}
\label{sec:introduction}
\fi
 \IEEEPARstart{G}{raphs} are
 widely adopted structures to model unique characteristics and complex relations among objects in a network, e.g., a social or citation network, or networks of financial transactions.
 This modeling potential can be leveraged by machine learning (ML) tasks to predict or recommend new or missing links, classify nodes into different groups, and detect fraudulent users or connections. However, the sparse, raw graph connectivity 
 does not readily lend itself to be used in ML tasks. 
 Instead, a regular, $d$-dimensional representation is more appropriate for learning. Graph embedding is the process of transforming a graph from its raw form to a latent $d$-dimensional representation, and it has received a surge of interest in the literature.\looseness=-1

The popularity of graph embedding made way for several techniques in the literature~\cite{verse18, LINE, deepwalk, node2vec}. However, most of these approaches do not scale to large-scale graphs. Even for small- and medium-scale graphs, \versex~\cite{verse18}, {\sc DeepWalk}~\cite{deepwalk}, \emph{node2vec}~\cite{node2vec}, and \linex~\cite{LINE} require hours of training. For instance, 
\versex
takes over three hours for a graph containing 4M vertices and 34M edges on 16 CPU cores. Various techniques have been proposed in the literature to make embedding faster. \harp~\cite{HARP} and \mile~\cite{mile18} 
proposed to shrink the graphs into smaller ones and applied a multi-level training procedure. However, due to the overhead of their heavy-weight coarsening, these approaches still fall short for large graphs. By using accelerators, such as GPUs, one can design and implement scalable graph embedding tools. To the best of our knowledge, the most successful GPU-based 
attempt is introduced by \graphvite~\cite{graphvite19}.

\looseness=-1 

In this paper, we present \malgo\footnote{
\url{https://github.com/SabanciParallelComputing/GOSH}}, a CPU-GPU parallel, fast, graph embedding algorithm for memory-restricted devices. \malgo applies a novel, parallel graph-coarsening that mitigates forming giant vertex sets. The coarsening algorithm iteratively shrinks the graph into multiple levels. Then a multi-level, unsupervised training is performed. The training starts at the coarsest level, and it continues with upper levels until the training on the original graph is completed, and the final embedding is obtained. \malgo is designed to handle large-scale graphs with a single GPU by utilizing a judiciously devised partitioning and scheduling algorithm. The contribution of this paper is threefold:\looseness=-1

\begin{itemize}[leftmargin=*]

    \item When the embedding does not fit into the GPU, state-of-the-art tools usually require multiple GPUs. On the contrary, \malgo applies a judicious partitioning and is faster than the multi-GPU tools thanks to its efficient, smart task scheduling, and its multi-level coarsening. Compared to its CPU-based counterpart, on graph {\tt com-orkut}, it is approximately three orders of magnitude faster while being more accurate. On the same graph, it is $36\times$ faster and scores $0.44\%$ higher AUCROC for link prediction compared to state-of-the-art. For node classification, it has comparable accuracy while being {$2.7\times$} faster.\looseness=-1 
    
    \item In the literature, coarsening has been applied by \mile~\cite{mile18} and \harp~\cite{HARP} to improve both the embedding performance and accuracy. Although the coarsening time 
    is negligible 
    for CPU-based embedding, this is not the case for GPUs. In this work, we propose a parallel and efficient coarsening algorithm that accelerates the embedding 
    by $73\times$ and improves the AUCROC by $2\%$ on {\tt com-orkut}.\looseness=-1
    
  
    \item 
    For different embedding dimensions, the best 
    GPU-implementation, that utilizes the device better, also differs. \malgo performs different 
    strategies to further increase the performance, especially for small $d$ values.\looseness=-1 
    
\end{itemize}

The paper is organized as follows: In Section~\ref{sec:not}, the notation and the essential background information are given. Section~\ref{sec:gosh} describes \malgo in detail including the techniques applied for link prediction and node classification. Section~\ref{s_coarsening} describes the parallel, multi-level coarsening algorithm, and Section~\ref{sec:meth_lg} describes our design and implementation proposed to handle large graphs. The experimental results are presented in Section~\ref{sec:exp}, and the related work is summarized in Section~\ref{sec:rel}. Section~\ref{sec:con} concludes the paper.\looseness=-1

\section{Notation and Background}
\label{sec:not}
A graph is a pair $G = (V, E)$ where $V$ is the set of vertices and $E \subseteq (V\times V)$ is the set of edges among them. Edges in undirected graphs are unordered vertex pairs, while in directed graphs, the order is significant. A $d$-dimensional embedding of $G$ is a $|V| \times d$ matrix ${\mathbf M}$. The vector $\mathbf{M}[v]$ is the embedding of vertex $v \in V$, and the value $\mathbf{M}[v][j]$ where $0 \leq j < d$ captures some information about $v$. Having a regular structure, the matrix $\mathbf{M}$ can be used in many ML tasks.\looseness=-1

There are many algorithms that embed a graph into a $d$-dimensional space. We model \malgo after \versex~\cite{verse18}, that is characterized by its small memory overhead and fast runtime, in addition to its ability to produce embeddings that reflect \emph{any} vertex-to-vertex similarity measure $Q$. \versex employs two distributions for each vertex $v \in V$ over all vertices. The first, $sim_{Q}^{v}$, is obtained from the empirical similarity between $v$ and $V \setminus \{v\}$ based on $Q$. The second, $sim_{E}^{v}$, is derived from the embedding matrix $\mathbf{M}$ using the cosine similarities between $v$'s embedding vector and those of $u \in V \setminus \{v\}$. \versex's objective function is minimizing the Kullback-Leibler (KL) divergence between these two distributions. In this paper, we choose $Q$ to be the adjacency similarity measure~\cite{verse18} that is defined as $Q(u, v)$ which is equal to $1$ for $(u, v) \in E$, and $0$, otherwise.


Training of the above objective, which
employs the {\em Noise Contrastive Estimation}~\cite{verse18}, is carried out by training a binary classifier to separate positive 
samples that are drawn from $sim_{Q}$ and negative
samples that are drawn from a noise distribution $N$, where the parameters of this classifier are the corresponding embedding vectors. More precisely, for a total of $e$ training epochs, we iterate through all 
$v \in V$ and process each 
$v$ by drawing a single positive 
sample $u$ from $sim_{Q}^v$, and $n_s$ negative 
samples $s_{1}, s_{2}, \cdots, s_{n_s}$ from $N$. Then, we minimize the negative log-likelihood of observing the positive sample and not observing the negative ones using a logistic regression model that updates the embedding vectors of $v$ and all the sampled vertices. Algorithm~\ref{alg:update} shows a single update of the source vertex $v$ and a (positive or negative) sample vertex $u$. In the algorithm, $b$ is a binary value that is equal to $1$ if the sample is positive, and $0$ if it is negative, $\odot$ is the dot-product operation, $\sigma$ is the {\tt sigmoid} function, and $lr$ is the classifier's learning rate. Table~\ref{tab:notation} summarizes the notation used in the paper. 

\renewcommand{\baselinestretch}{0.95}
\newcommand{\updateembedding}{{\sc UpdateEmbed}}
\begin{algorithm}
\KwData{$\mathbf{M}[v]$, $\mathbf{M}[sample]$, $b$, 
        $lr$}
 \KwResult{$\mathbf{M}[v]$, $\mathbf{M}[sample]$}
 $score \leftarrow b - \sigma(\mathbf{M}[v] \odot \mathbf{M}[sample]) \times lr$ \;
 $\mathbf{M}[v] \leftarrow \mathbf{M}[v] + \mathbf{M}[sample] \cdot score$\;
 $\mathbf{M}[sample] \leftarrow \mathbf{M}[sample] + \mathbf{M}[v] \cdot score$\;
\caption{\updateembedding}\label{alg:update}
\end{algorithm}
\renewcommand{\baselinestretch}{1}

\begin{table}
\centering
  \caption{\small{Notation used in the paper.}}
  \label{tab:notation}
  \scalebox{0.98}{
  \begin{tabular}{ll}
    \hline
    \textbf{Symbol} & \textbf{Definition}\\
    \hline
    $G_0 = (V_0, E_0)$ & The original graph to be embedded.\\
    $G_{i} = (V_{i}, E_{i})$ & Represents a graph, which is coarsened $i$ times.\\
    $\Gamma(u)$ & Neighborhood of $u$.\\\hline
    $d$ & $\#$ vertex features, i.e., the embeddıng dimension.\\
    $n_s$ & $\#$ negative samples per vertex.\\
    $\sigma$ & Sigmoid function.\\
    $e$ & Total number of epochs that will be performed\\
    $lr$ & Learning rate.\\\hline
    $D$ & Total amount of coarsening levels. \\
    ${\mathcal G}$ & The set of coarsened graphs created from $G_0$. \\
    $p$ & Smoothing ratio for epoch distribution.\\
    $e_i$ & $\#$ epochs for coarsening level $i$.\\
    ${\mathbf M}_{i}$ & Embedding matrix obtained for $G_{i}$.\\
    ${\mathcal M}$ & The set of mappings used in coarsening.\\
    $map_i$ & Mapping information from $G_{i-1}$ to $G_{i}$. \\\hline
    $K$ & $\#$ parts in ${\mathcal V}$. \\
    ${P}_{GPU}$ & $\#$ embedding parts to be placed on the GPU. \\
    $S_{GPU}$ & $\#$ sample pools to be placed on the GPU. \\
    $B$ & $\#$ positive samples per vertex in a  sample pool. \\
    $\mathbf{K}_{i, j}$ & embedding kernel for sub-graphs $i$ and $j$. \\
  \hline
\end{tabular}}
\end{table}

\section{\malgo: Graph Embedding on a GPU}
\label{sec:gosh}
To describe \malgo, we begin with a high-level explanation of the tool, followed by the GPU acceleration details in Section~\ref{sec:gpu_imp}. Section~\ref{sec:walks} explains the random-walk-based version of \malgo 
to perform embedding for node classification. The coarsening algorithm and the techniques applied to handle large graphs are 
described in Sections~\ref{s_coarsening}~and~\ref{sec:meth_lg}.


\begin{figure*}
    \centering
    \includegraphics[width=0.90\textwidth]{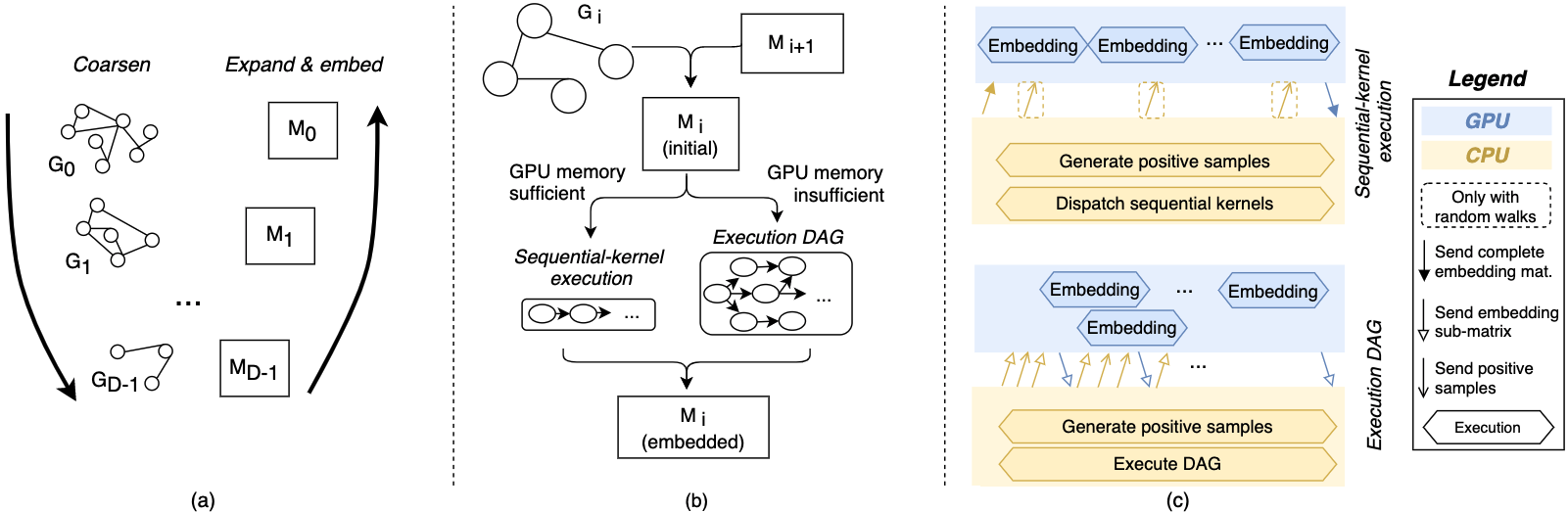}
    \caption{{\small{An overview of \malgo: (a) The input graph, $G_0$, is iteratively coarsened into smaller graphs until an exit condition is met. Then, starting from the smallest, $G_{D-1}$, each graph $G_{i}$ is embedded. 
    (b) To embed $G_{i}$, the embedding $M_{i+1}$ is projected onto $G_{i}$ to initialize $M_{i}$ which is then fine tuned. If $G_{i}$ and $M_{i}$ fits to the GPU memory \malgo carries out \textit{Sequential-Kernel Execution}~(Section~\ref{sec:gpu_imp}). Otherwise, it carries out \textit{Directed Acyclic Graph Execution}~(Section~\ref{sec:meth_lg}). (c) In \textit{Sequential-Kernel Execution}, CPU dispatches kernels to GPU. Meanwhile, GPU executes the kernels sequentially. If random-walk sampling is used CPU generates positive samples and sends them to GPU. On the other hand, in DAG execution, the CPU continuously generates positive samples, sends them to GPU, and exchanges submatrices with it. Meanwhile, the GPU concurrently performs embedding.}}}
    \label{fig:overview}
    \vspace*{-2ex}
\end{figure*}

Algorithm~\ref{alg:malgo}, the main body of \malgo, takes a graph $G_{0}$ as input and computes the embedding $\mathbf{M}_{0}$. The embedding is performed in two stages:
\begin{enumerate}[leftmargin=*]
     \item creating a set ${\mathcal G} = \{G_0, G_1, \ldots, G_{D-1}\}$ of graphs coarsened iteratively starting from $G_{0}$ as in the left of Fig.~\ref{fig:overview}.a, where one or more nodes in $G_{i-1}$ are uniquely represented by a super node in $G_{i}$ (Line~\ref{ln:malgo:coarsening} of Alg.~\ref{alg:malgo}),
     \item starting from the coarsest graph, $G_{D-1}$, training the embedding $\mathbf{M}_{i}$ of graph $G_{i}$ and projecting it to $\mathbf{M}_{i-1}$, to later train 
     it as shown in 
     Fig.~\ref{fig:overview}.a~(Lines~\ref{ln:malgo:bfor}--\ref{ln:malgo:efor} of Alg.~\ref{alg:malgo}). 
\end{enumerate}
\noindent The algorithm stops once ${\mathbf M}_0$ is obtained. We obtain ${\mathbf M}_{i-1}$ from ${\mathbf M}_{i}$ by using the mapping information for ${\mathbf G}_{i-1}$, where $M_{i}[u] = M_{i-1}[v]$ iff $u \in V_{i}$ is a super node of $v \in V_{i-1}$.\looseness=-1

\renewcommand{\baselinestretch}{0.95}
\begin{algorithm}
\KwData{$G_{0}$, $n_s$, $lr$, 
        $lr_d$, 
        $p$, $e$, $threshold$, $P_{GPU}$, $S_{GPU}$, $B$}
 \KwResult{${\mathbf M}$}
 ${\mathcal G} \leftarrow $\coa$(G_{0}, threshold)$\;\label{ln:malgo:coarsening}
  Randomly initialize ${\mathbf M}_{D-1}$\;
 \For{$i$ from $D-1$ to $1$}
 {\label{ln:malgo:bfor}
    $e_i \leftarrow$ {\sc calculateEpochs}($e$, $p$, $i$)\; 
    \If{$G_{i}$ and ${\mathbf M}_{i}$ fits into GPU}
    {\label{ln:malgo:ifstatment}
        {\sc CopyToDevice}($G_{i}$, ${\mathbf M}_{i}$)\;\label{ln:malgo:copy}
        ${\mathbf M}_{i} \leftarrow$ \smalltrain($G_{i}$, ${\mathbf M}_i$, $n_s$, $lr$, $lr_d$, $e_i$)\;\label{ln:malgo:smalltrain}
    }
    \Else
    {
         ${\mathbf M}_{i} \leftarrow$ \lge($G_{i}$, ${\mathbf M}_i$, $n_s$, $lr$, \\\hspace*{19ex}$lr_d$, $e_i$,$P_{GPU}$, $S_{GPU}$, $B$)\;\label{ln:malgo:bg}
    }
        ${\mathbf M}_{i-1} \leftarrow$ {\sc ExpandEmbedding}(${\mathbf M}_{i}$, $map_{i-1}$)\;\label{ln:malgo:expand}

}\label{ln:malgo:efor}
{\bf return $\mathbf{M}_0$}\;
\caption{\malgo}\label{alg:malgo}
\end{algorithm}
\renewcommand{\baselinestretch}{1}

\malgo is able to accelerate the embedding of large-scale graphs whose memory footprint exceeds the capabilities of contemporary GPUs. Real-world graphs can have hundreds of millions of vertices; for $|V| = 128M$, the embedding matrix $\mathbf{M}$ with $d = 128$ takes $64$GB in single precision and $128$GB in double precision. 
For each $G_{i} \in \mathcal{G}$, \malgo checks whether the GPU can store both $G_{i}$ and $\mathbf{M}_{i}$ (Line~\ref{ln:malgo:ifstatment}). If this is the case, they are both copied to the GPU, and the training proceeds in a single step~(Lines~\ref{ln:malgo:copy}-\ref{ln:malgo:smalltrain}). Otherwise, the {\em large-graph embedding} schema described in Section~\ref{sec:meth_lg} is used, in which $\mathbf{M}_{i}$ is partitioned into sub-parts to be copied and processed on the GPU. In this schema, the samples are generated on the host and dynamically sent to the GPU.\looseness=-1

Multilevel coarsening introduces a peculiar work distribution problem. Let $e$ be the total number of training epochs performed on all levels; one can simply evenly distribute them. 
However, if more epochs are reserved for the coarser levels, the overall embedding becomes faster. Besides, training on coarser levels significantly impacts the final embedding since the updates are projected to many finer-level vertices. On the other hand, more epochs for the higher levels potentially make the final embedding more fine-tuned. Therefore, one must {\em distribute the epoch budget} cleverly. \malgo employs a mixed strategy; 
we distribute a portion $p$ of the epochs uniformly across all levels and distribute the remaining $(1-p)$ portion of epochs geometrically. More formally, the $i$th level training uses $e_i = e/D + e'_i$ epochs where $e'_i$ is half of $e'_{i+1}$. We leave the value $p$, dubbed the {\em smoothing ratio}, as a configurable parameter to allow an interplay between accuracy and performance. The learning rate is another important parameter that has a significant impact on embedding quality. \malgo uses the same initial learning rate $lr$ for all levels but decreases it linearly in each epoch. More formally, at training level $i$, in epoch $j$, the learning rate is equal to $lr \times \max\left(1 - \frac{j}{e_i}, 10^{-4}\right)$.

\subsection{Graph embedding in \malgo} \label{sec:gpu_imp}
\malgo implements a GPU-accelerated, lock-free training step, and similar to \versex, it uses SGD-based optimization. As shown in~\cite{hogwild}, a lock-free SGD implementation which does not take {\em race conditions}, i.e., simultaneous updates, into account does not negatively impact the learning quality on multi-core CPUs significantly. However, on a GPU with millions of parallel threads, {especially with coarsening}, we observed that race conditions deteriorate the quality. To mitigate the impact of race conditions, we serialize the training epochs, i.e., no two training epochs are executed in parallel. {Section~\ref{sec:ep_sync} describes this effect in more detail.}

In an epoch for ${\mathbf M}_{i}$, \malgo traverses $V_{i}$ and assigns a single source vertex to a single GPU warp. For each source vertex, multiple positive and negative samples are processed sequentially by the warp assigned to that vertex, and updates are performed as in Algorithm~\ref{alg:update}. With this vertex-to-warp mapping, no vertex $v \in V_{i}$ can be a source vertex for two concurrent updates, but $v$ may be (positively or negatively) sampled by one or more vertices. Furthermore, its assigned warp can also be active. 
Therefore, reads and writes on $\mathbf{M}_i$ are not completely race free. However, our experiments show that synchronizing epochs in this way is sufficient to perform the parallel embedding process robustly.\looseness=-1 

The positive and negative samples for source vertices are generated on the GPU as shown in Algorithm~\ref{alg:EA}. A positive sample for source $v \in V_{i}$, is a vertex $u \in V_{i}$ s.t. $u \in \Gamma_{G_{i}}(v)$.
As mentioned in Section~\ref{sec:not}, the negative samples are drawn from a noise distribution, which we model as a uniform random distribution over $V_{i}$ similar to~\cite{verse18}. 

\renewcommand{\baselinestretch}{0.95}
\begin{algorithm}[htbp]
 \KwData{$G_{i}$, ${\mathbf M}_i$, $n_s$, $lr$, $e_i$}
 \KwResult{${\mathbf M}_{i}$}
  \For{$j = 0$ to $e_i - 1$}{ 
  $lr' \leftarrow lr \times max \left(1 - \frac{j}{e_i}, 10^{-4}\right)$\;
  \tcc{Each $src$ is assigned to a warp}
  \For{$\forall src \in V_i$ \bf in parallel} {
    $u \leftarrow$ {\sc GetPositiveSample}($G_i$)\;
     \updateembedding($\mathbf{M}_i[src]$, ${\mathbf M}_i[u]$, $1$, $lr'$)\; 
      \For{$k = 1$ to $n_s$}{
         $u \leftarrow$ {\sc GetNegativeSample}($G_i$)\;
         \updateembedding($\mathbf{M}_i[src]$, ${\mathbf M}_i[u]$, $0$, $lr'$)\;
       }
    }
 }
\caption{\smalltrain}\label{alg:EA}
\end{algorithm}
\renewcommand{\baselinestretch}{1}

When a warp carries out the updates for a source vertex $src$, its threads perform $(1+n_s) \times d$ accesses to $\mathbf{M}_{i}[src]$. Therefore, larger values of $n_s$ and $d$ increase global memory accesses and deteriorates performance dramatically. To counter this issue, before $src$ is processed, $\mathbf{M}_{i}[src]$ is copied from the global memory of the GPU to the shared memory of a streaming multiprocessor. This way, for all consecutive positive and negative updates, reads and writes to $\mathbf{M}_{i}[src]$ are performed on the shared memory. After all the updates are executed, ${\mathbf M}_{i}[src]$ is copied back to global memory. For a sampled vertex $u$, on the other hand, $\mathbf{M}_{i}[u]$ is always kept in global memory since it is only read and written once. Accesses to $\mathbf{M_{i}}[u]$ for a vertex $u$ are coalesced, where a warp's threads perform the accesses in a round-robin fashion. More precisely, ${\mathbf M}_{i}[u][j + (32 \times k)]$ is accessed by thread $j$ on its $k$th access where $32$ is the number of threads within a warp.\looseness=-1

\subsubsection{Embedding for small dimensions} \label{sec:small_dimensions}
When embedding with $d \leq 16$ dimensions and with 32 threads in a warp, a single source vertex does not occupy all the threads in a warp. When $d$ is small, since $32-d$ threads in each warp remain idle, the device will be underutilized. For this reason, \malgo employs a specialized vertex-warp distribution for small $d$. Depending on the dimension, we assign $2$ or $4$ source vertices to a single warp, and hence, $16$ or $8$ threads, respectively, are assigned to a single source. 

\subsection{Random-walk-based sampling for classification}\label{sec:walks}

Our preliminary experiments show that  \versex-based sampling  can not compete with the state-of-the-art on node classification. Instead, we employ {\em random-walk-based sampling}, which is a common technique in embedding literature, and implement \malgow which samples from random walks on the host machine and sends them to the GPU in an online fashion. To sample positive edges for node classification, we start a random walk of length $\ell$ and sample every pair of vertices within a distance of $g$.\looseness=-1

We opted to carry out the random walks and sampling on the CPU for three reasons. First, \malgo now utilizes CPU and GPU at the same time. Furthermore, a single walk of length $\ell$ and sampling distance $g$ require storing a minimum of $g$ elements. By sampling on the host, we do not impose the burden on GPU memory, which is already the main bottleneck.
Second, we want to exploit the relative efficiency of CPUs over GPUs on irregular memory accesses, which happen due to the nature of random walks and the graph sparsity. Last, Zhu~et~al. show that
shuffling the samples before using them 
produces better results in downstream ML tasks~\cite{graphvite19}. 
Acquiring the samples from walks on the host allows them to be shuffled more easily.

\section{Coarsening Graphs for Fast Embedding}\label{s_coarsening}

\malgo coarsens a graph $G_i = (V_i, E_i)$ while retaining the structure of the coarse graph and maximizing coarsening {\em efficiency} and {\em effectiveness}. We measure the $i$th level efficiency by the rate of shrinking, ${(|V_{i - 1}| - |V_{i}|)}/{|V_{i - 1}|}$. On the other hand, the effectiveness is measured in terms of embedding quality. Our approach, \coa, is agglomerative and generates the clusters similar  to~\cite{HARP}. The vertices in $V_{i}$ are processed one by one. When processing $v \in V_{i}$, if it is not yet marked it is marked and mapped to a cluster, i.e., a new {\em super} vertex in $V_{i+1}$, and its edges are processed. If an edge $(u, v) \in E_{i}$ with an unmarked $u$ exists, $u$ is added to $v$'s cluster. Finally, all the vertices in $v$'s cluster are shrunk into a super vertex $v_{sup} \in G_{i+1}$.   

\coa can preserve both the first- and second-order proximities of vertices as defined in~\cite{LINE}. The former proximity measures the pairwise connection between two vertices, while the latter measures the similarity between vertices' neighborhoods. It accomplishes that by collapsing vertices belonging to the same neighborhood around a local hub vertex. However, we observed that a naive implementation usually merges two giant, {\em hub} vertices. This degrades the effectiveness and efficiency of the coarsening. The effectiveness degrades since the coarsened graphs are not structurally similar in lower coarsening levels because a small number of giant super vertices represent almost all the vertices on the finer graphs. Having a small set of giant super vertices also inhibits coarsening the graph further, degrading the efficiency. To delay and if possible, avoid such cases, we introduce a new merging condition, {\bf hub$^2$-restriction}, where $u \in V_{i}$ is not placed into the cluster of $v \in V_i$ if $|\Gamma_{G_{i}}(u)|$ and $|\Gamma_{G_{i}}(v)|$ are both larger than $\frac{|E_{i}|}{|V_{i}|}$. This way, two hub vertices whose degrees are greater than the density of $G_{i}$ will no longer be in the same cluster. Our experiments show that this simple rule significantly improves both coarsening effectiveness and efficiency.\looseness=-1

When a vertex is marked and added to a cluster, its neighbors are not processed and do not contribute to the coarsening. If vertices are processed in an arbitrary order, the efficiency may degrade since vertices with small neighborhoods can lock large vertices. Hence, when an edge $(u, v) \in E_{i}$ is used for coarsening via a hub-vertex $v \in V_{i}$, we prefer $u$ to be inserted into $v$'s cluster to maximize efficiency. We achieve this by a new heuristic, dubbed {\bf ordering}, resulting in substantial improvements in coarsening efficiency. With {\bf ordering}, the vertices are ordered to process the vertices with larger neighborhoods earlier.

The coarsening process is shown in Alg.~\ref{alg:MEC}. The algorithm takes an input graph $G = G_{0}$ and returns a set ${\mathcal G}$  of graphs and the mapping information that will be used to project the embedding matrices $\mathcal{M}$. We initialize ${\mathcal G}$ and ${\mathcal M}$ as $\{G_{0}\}$ and $\emptyset$, respectively. Coarsening starts with $i = 0$ and terminates either when a graph $G_{i+1}$ with a vertex count that is less than $threshold$ is obtained or $|V_{i+1}|$ is larger than $80\%$ of $|V_{i}|$. {We observed that $80\%$ works well for scale-free graphs, however it is experimentally tuned and left as a parameter.} Additionally, we store the mapping information $map_i$ which is used to shrink $G_i$ to $G_{i+1}$, and later use it to project the embedding matrix $M_{i+1}$ and initialize $M_{i}$. \malgo uses  $threshold = 100$ for all the experiments.

\renewcommand{\baselinestretch}{0.95}
\begin{algorithm}[htbp]
 \KwData{$G_0 = (V_0, E_0)$, $threshold$}
 \KwResult{${\mathcal G}$, ${\mathcal M}$}
 ${\mathcal G} \leftarrow \{G_{0}\}$, ${\mathcal M} \leftarrow \emptyset$, $i \leftarrow 0$\;
 \Do{$|V_{i}| > threshold$ and $|V_{i}| \leq |V_{i-1}|\times 0.80$}
    {
    $order \leftarrow$ {\sc Sort}($G_{i}$)\;\label{ln:sort}
    \lFor{$v \in V_i$} {$map_i[v] \leftarrow - 1$}
    $\delta \leftarrow |E_i| / |V_i|$\;
    $cluster \leftarrow 0$\;
    \For{$v$ in $order$} { \label{ln:fors}
        \If{$map_i[v] = -1$} {
            $map_i[v] \leftarrow cluster$\;\label{ln:map1}
            $cluster \leftarrow cluster + 1$\;
            \ForEach{$(v, u) \in E_i$} {
                    \If{$|\Gamma_{G_i}(v)| \leq \delta$ or $|\Gamma_{G_i}(u)| \leq \delta$} {
                                    \If{$map_i[u] = -1$} {
                        $map_i[u] \leftarrow map_i[v]$\;\label{ln:map2}
                    }
                }
            } 
        }
    }\label{ln:fore}
    $G_{i+1} \leftarrow$ {\sc Coarsen}($G_{i}, map_{i}$)\;\label{ln:coarse}
    ${\mathcal G} \leftarrow {\mathcal G} \cup \{G_{i+1}\}$, ${\mathcal M} \leftarrow {\mathcal M} \cup \{map_{i}\}$, $i \leftarrow i + 1$\;
 }
 \caption{\coa}\label{alg:MEC}
\end{algorithm}
\renewcommand{\baselinestretch}{1}

\subsection{Complexity analysis}
\malgo employs the {\em Compressed Storage by Rows}~(CSR) graph data structure which stores the vertex neighborhoods in an adjacent manner. 
There are three stages in \coa; sorting~(Line \ref{ln:sort}), mapping~(Lines \ref{ln:fors}--\ref{ln:fore}) and coarsening~(Line \ref{ln:coarse}). The sorting stage uses \emph{counting sort} whose time complexity is $\mathcal{O}(|V| + |E|)$. For mapping, the algorithm must traverse all the edges, leading to a time complexity of $\mathcal{O}(|V| + |E|)$. Finally, coarsening the graph requires the vertices to be sorted with respect to their mappings and then going through all the vertices and their edges within the CSR, incurring a time complexity of $\mathcal O(|V| + |E|)$.\looseness=-1



\subsection{Parallelization of \malgo coarsening}\label{s_coarsening_parallel}
On a CPU, the embedding process dominates the total time.
However, this is not the case when a fast, GPU-based embedding is employed, such as the one in \malgo. Then 
the coarsening overhead can no longer be neglected. Hence, a parallel coarsening is a must for \malgo.

For coarsening, traversing $V_{i}$ in parallel and 
mapping with no synchronization can yield inconsistencies. We avoid the race conditions by using a single lock per each $map_i$. While updating $map_{i}[v]$ and $map_{i}[u]$ as in Lines~\ref{ln:map1}--\ref{ln:map2} of Alg.~\ref{alg:MEC}, a coarsening thread tries to lock both $map_i[v]$ and $map_i[u]$. If the locks are acquired, the thread continues.
Otherwise, it skips the current candidate and continues with the next vertex. Besides, to avoid race conditions on the counter $cluster$, the algorithm uses the hub-vertex id for mapping. More precisely, $map_i[v]$ is set to $v$ as opposed to Line~\ref{ln:map1} in the sequential algorithm. It should be noted that with the parallel implementation, $map_i$ does not provide a mapping to vertex IDs in $G_{i+1}$. This can be corrected in $\mathcal{O}(|V|)$ time through a sequential traversal of $map_i$, which first detects/counts the vertices that have $map_i[v] = v$ and resets the values of $map_i$ for all the vertices.\looseness=-1

Constructing the coarsened graph in parallel is not easy since after the mapping is generated, we do not yet know the degrees of the (super) vertices in $G_{i+1}$. To mitigate this issue, a private region of memory $E^j_{i+1}$ is allocated for each thread $t_j$ where $1 \leq j \leq \tau$. Each thread creates edge lists of the new vertices in these private regions, which are merged on a different location with size $|E_{i+1}|$. This happens via a sequential scan to find the region in $E_{i+1}$ for each thread's private information, and copying that private information to $E_{i+1}$. An important issue with this approach is load imbalance. A static vertex-to-thread assignment can reduce embedding performance since the degree distribution of the original graph can be skewed and more skewed for the coarsened graphs. For this reason, a dynamic scheduling strategy that uses small batch sizes is used in all the steps above.\looseness=-1

\section{Embedding Large Graphs on a GPU}
\label{sec:meth_lg}
\malgo can bypass the memory limitations of the GPU. Unlike other GPU-based approaches such as~\cite{graphvite19, edges}, whose hardware requirements increase with the memory requirements, \malgo requires only a single GPU to accelerate the embedding of any arbitrarily large graph. The embedding algorithm adopted in this work so far requires storing the embedding matrix $\mathbf{M}$ as well as the graph on the GPU, whose memory complexities are $\mathcal{O}(d \times |V|)$ and $\mathcal{O}(|V| + |E|)$, respectively. For a graph with $10^9$ vertices and $d = 10^2$, this is more than any modern GPU can provide. Here we describe a novel partitioning schema and DAG execution model, which can be used when the memory requirement of graph embedding exceeds the GPU memory size.\looseness=-1

Instead of storing $\mathbf{M}$ as a whole on the GPU, we employ a partitioning schema as in~\cite{graphvite19, edges, pbg19} which partitions the vertex set $|V|$ into $K$ disjoint sets $V^{0}, V^{1}, \cdots,$ $V^{K-1}$ each with a corresponding submatrix ${\mathbf M}^{0},$ ${\mathbf M}^{1},$ $\cdots, {\mathbf M}^{K-1}$. These submatrices are copied in and out of the GPU to be embedded. As shown in Alg.~\ref{alg:EA}, a single update accesses no more than two embedding vectors. Hence, the correctness of an embedding requires that for every edge $(u, v) \in E$, the embedding vectors of the vertices $u, v \in V$ must be simultaneously on the GPU at some time point. For that reason, \malgo splits the embedding procedure into rounds where each round is processed by an embedding kernel ${\bf K}_{i,j}$ on a vertex-set pair $(V^{i}, V^{j})$ where $0\leq i \leq j < K$. For each vertex $u \in V^{i}$~(and if $i \neq j$, for each vertex $v \in V^{j}$), ${\bf K}_{i,j}$ performs up to $B$ positive updates, and for each positive update, $s$ negative updates. The positive samples are generated on the host in parallel and sent to the GPU as they are needed, as opposed to the base algorithm, in which they are generated on the GPU. This removes the need to store the graph on the GPU, and concurrently utilizes the host's resources. The negative samples are generated on the GPU by sampling randomly from the opposite embedding matrix. This execution schema implies that in a single embedding round, \textit{at most} $B \times K$ positive samples are executed \textit{per vertex}. For a vertex $v \in V^{i}$, \textit{no} updates will be carried out by a ${\bf K}_{i, j}$ or ${\bf K}_{j, i}$ if no neighbor of $v$ appears in $V_{j}$. In this schema, we run $r = \frac{e}{B}$ rounds and use an atomic global counter to keep the number of samples from exceeding $e\times|V|$.\looseness=-1 

\subsection{Kernel execution order}

\begin{figure}[h!]
\hspace*{-10ex}
    \centering
    \begin{subfigure}[b]{0.35\linewidth}
        \centering
        \includegraphics[width=0.95\linewidth]{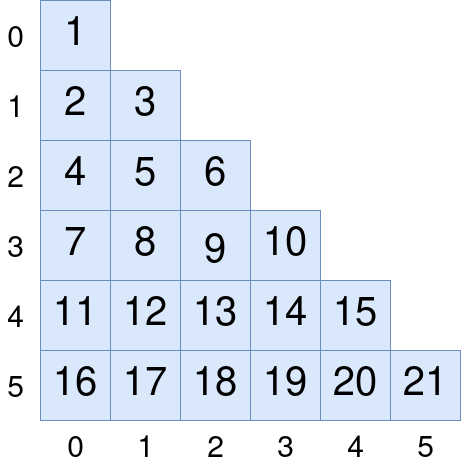}
        \caption{}
        \label{fig:bg:order}
    \end{subfigure}    \begin{subfigure}[b]{0.45\linewidth}
        \centering
        \includegraphics[width=1.40\linewidth]{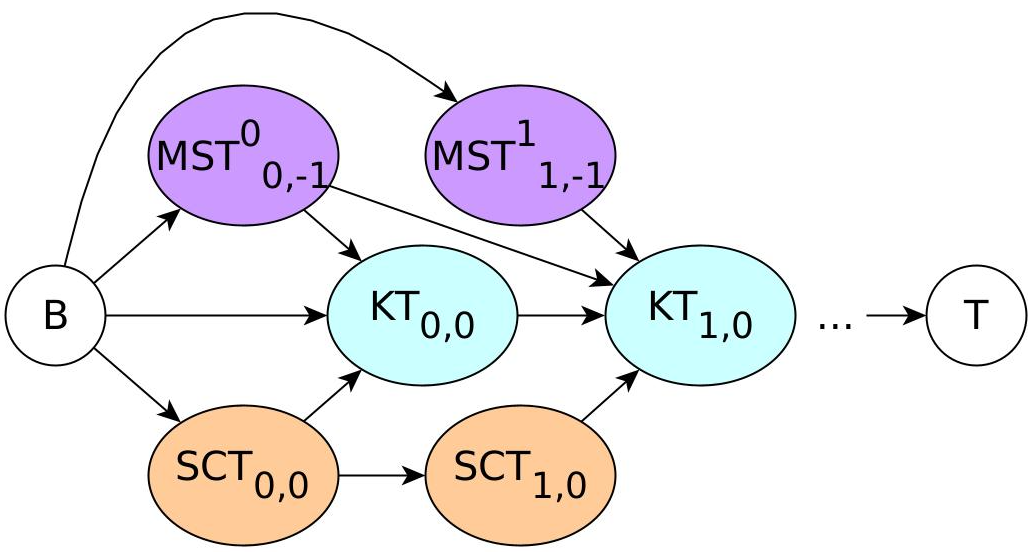}
        \caption{}
        \label{fig:bg:dag}
    \end{subfigure}
    \label{fig:graph}
    \caption{\small{(a) Partitioning of $V$ into 6 and kernel order: each box at row $i$ and column $j$ denotes  $\mathbf{K}_{i,j}$ and the numbers denote the order. (b) A DAG with the tasks grouped into 3
``lanes''; (1) submatrix swap, (2) sample copy, and (3) kernel tasks.}}
\vspace{-4ex}
\end{figure}

The order in which kernels are executed in an embedding has a direct impact on the embedding efficiency. Although it does not change the amount of updates performed, the order directly affects the number of  submatrix transfers happening between the host and the GPU. 
We use the {\em inside-out ordering}~\cite{pbg19} as it minimizes the number of submatrix copies to and from the GPU and produces high quality embeddings. We define the part pairs on which the kernels  operate as $\mathcal{X} = (V^{a_{0}}, V^{b_{0}}), (V^{a_{1}}, V^{b_{1}}), \cdots, (V^{a_{\ell}}, V^{b_{\ell}})$ where $\ell = \frac{K(K+1)}{2}$ and
\[(V^{a_j}, V^{b_j}) = 
\begin{cases}
(V^{0}, V^{0}) & $j = 0$ \\\
(V^{a_{j-1}}, V^{b_{j-1}+1}) & j > 0 \text{ and } a_{j-1} > b_{j-1}  \\
(V^{a_{j-1}+1}, V^{0}) & a_{j-1} = b_{j-1} \\
\end{cases}\]
A toy example of a kernel order for a graph with six parts is shown in Fig.~\ref{fig:bg:order}.

\subsection{Number of submatrix and sample pool bins } 
While partitioning the embedding matrix, there are two important hyper-parameters; (1) the number of submatrices $P_{GPU}$, and (2) the number of sample pools $S_{GPU}$ simultaneously stored on the GPU. For correctness, we require that $P_{GPU} \geq 2$. Although fewer submatrices are easier to handle, more would allow more overlap between memory transfers and computation. For example, assume that ${\mathbf M}^{0}$, ${\mathbf M}^{1}$ and ${\mathbf M}^{5}$ are currently on the GPU, and the three upcoming kernels are $\mathbf{K}_{5,0}$, $\mathbf{K}_{5,1}$, and $\mathbf{K}_{5,2}$. After $\mathbf{K}_{5,0}$ is done, while $\mathbf{K}_{5,1}$ is running, $\mathbf{M}_{0}$ can be switched out and replaced with $\mathbf{M}_{2}$. Increasing $P_{GPU}$ also increases $K$ and decreases the number of vertices in a single vertex set. Hence, more memory copies would be required in a single round. We empirically find that $P_{GPU} = 3$ provides the best performance.\looseness=-1

As mentioned before, positive samples are generated on the host and sent to the GPU as they are needed. We allocate $S_{GPU}$ sample pool bins on the GPU such that each pool can store the positive samples used by a single kernel $\mathbf{K}_{i,j}$. Having multiple pools on the GPU allows further concurrency between kernels. However, it also incurs a memory overhead reducing the available memory size for embedding submatrices.
We find that $S_{GPU} = 2$ exposes enough concurrency without increasing $K$ too much.
\newcommand{\samplemanager}{{\tt SampleManager}\xspace}
\newcommand{\poolmanager}{{\tt PoolManager}\xspace}
\newcommand{\switchsubmatrices}{{\sc SwitchSubMatrices}\xspace}
\newcommand{\embeddingkernel}{{\sc EmbeddingKernel}\xspace}
\newcommand{\getnextsubmatrix}{{\sc NextSubMatrix}\xspace}

\subsection{Directed acyclic graph execution model}\label{ch:dag}
Embedding a graph using the proposed partitioning framework involves various parallel task executions and a myriad of dependencies among them. The correctness of the algorithm can be maintained using global synchronization, but this can potentially degrade the performance. Instead, we adopt a {\em Directed Acyclic Graph}~(DAG) execution model to resolve dependencies and further optimize the algorithm. {In this model, a task must only synchronize with its incoming neighbors}. We first define three task types running on the host that encompass the embedding:
\begin{enumerate}
    \item {\bf Submatrix-swap task} ($MST_{i,j}^{k}$): dispatches a GPU-to-host transfer of  submatrix ${\mathbf M}_{i}$ out of the $k$th submatrix bin, where $k < P_{GPU}$. It also dispatches a host-to GPU transfer of ${\mathbf M}_{j}$ to the same bin.
    \item {\bf Sample-pool-copy task} ($SCT_{i,j}$): dispatches a memory copy of the sample pool $\mathbf{S}_{i,j}$ to the GPU. {The exact bin on the GPU 
    is resolved at runtime}. 
    \item {\bf Kernel-execution task} ($KT_{i,j}$): dispatches a  kernel $\mathbf{K}_{i,j}$ to the GPU. 
\end{enumerate} 
We construct an embedding DAG by associating these tasks and modeling the dependencies among them as directed edges. We associate an execution queue with the DAG to hold the {\em ready-to-be-executed} nodes whose incoming neighbors are completed. This queue uses a team of threads to execute the tasks in parallel. 

{When a task is executed, it asynchronously dispatches the associated GPU job and terminates without waiting for its execution. 
Such a loose coupling reduces the GPU-host synchronization overhead, reduces the host workload, and saves more resources for the far more expensive sampling. 
We used native GPU constructs, {\tt cudaStreams} and {\tt cudaEvents}, to realize the DAG dependencies on the GPU. {\tt cudaStreams} are used to create sequences of synchronous GPU jobs. Since the DAG cannot be partitioned into linear sequences/streams, we   set inter-stream dependencies via {\tt cudaEvents} among the jobs on different streams.}

{\malgo allocates a {\tt cudaStream} for every sample pool- and submatrix-bin on the GPU since copy operations on the same bin require a natural sequential dependency. Also, it allocates two {\tt cudaStreams} for embedding kernels 
to exploit their potential overlap. 
Each task is associated 
with a {\tt cudaEvent} that is triggered at the end of the dispatched GPU job. 
When a task dispatches a GPU job, it also instructs it to wait for the {\tt cudaEvents} of its incoming neighbors' GPU jobs, and this makes the dependencies of the GPU completely isolated from the host's. The only exceptions to 
this are the dependencies among kernel-execution tasks; we want the embedding kernels to overlap whenever possible.}

Overall, the task dependencies are summarized as follows: the task $KT_{i,j}$ depends on the tasks which bring its sample pool and submatrices to the GPU, i.e., $SCT_{i,j}$, $MST_{\ast, i}^{\ast}$, and $MST_{\ast, j}^{\ast}$. The tasks $KT_{i,j}$ and $SCT_{i,j}$ depend on $KT_{m,n}$ and $SCT_{m,n}$, respectively, where $(m, n)$ is the pair preceding $(i, j)$ in the kernel order~(the dependency between $KT_{m,n}$ and $KT_{i, j}$ does not involve {\tt cudaEvents} as described before). Finally, the task $M_{i, j}^{k}$ depends on all $KT$s which use the submatrix $i$, stored in submatrix bin $k$. A sample DAG is shown in Fig.~\ref{fig:bg:dag}.\looseness=-1

{
Unlike submatrix copy tasks, sample pool tasks do not know the exact copy bin on the GPU. 
This is a deliberate design choice made since the processing times of sample pools can vary, and dynamically allocating the GPU pool bins to the first available sample-pool-copy task improves the utilization of bins. To implement this, we leverage 
mutex-protected 
host variables that are 
used by the sample-pool-copy and kernel-execution tasks to communicate.}

\section{Experimental Results}
\label{sec:exp}

The embedding quality of \malgo, \versex, \mile, \pbg, and \graphvite are evaluated with link prediction and node classification tasks, which are widely used for evaluation in the literature~\cite{verse18, node2vec, graphvite19, pbg19}.

\subsection{Details of the experiments}
The details, e.g., datasets and their use for ML tasks, hardware, and the baseline tools are given below.\\

\noindent{\bf{\underline{Hardware specifications}}}:
We run our experiments on three servers. The first, {\bf Arch-1}, is equipped with an Nvidia V100 GPU with 16GB memory, 348GB RAM, and 2 Intel Xeon Gold 6148 processors, each with 20 cores running at 2.40GHz.
The second, {\bf Arch-2}, uses a single Titan X Pascal GPU with 12GB of global memory, 198GB RAM, and two Intel Xeon E5-2620 v4 processors, each with 8 CPU cores running at 2.10 GHz. Lastly, {\bf Arch-3} has $4$ Nvidia A100 GPUs with $80$GB of memory, $64$ cores on an AMD 7742 CPU that runs on 2.24GHz, and $512$GB of memory.   

{The servers have {\tt CentOS 7.3.1611 (Core)} and {\tt Ubuntu 4.4.0-159}, and {\tt CentOs 7.9} as operating systems, respectively. All the CPU codes are compiled with {\tt gcc 7.0.1} with {\tt -O3} as the optimization parameter. For CPU parallelization, {\tt OpenMP} multithreading is used in general. The GPU codes are compiled with {\tt nvcc} with {\tt CUDA 10.1} and {\tt -O3} optimization flag. The GPU is connected to the server via {\tt PCIe 3.0 x16}. For GPU implementations, all the relevant data structures are stored on the GPU memory.}\looseness=-1

\noindent{\bf{\underline{Datasets used for the experiments}}}:
 We use six {\em medium-scale} and three {\em large-scale} graphs to fairly and thoroughly evaluate the tools. We choose graphs that differ in terms of their origin, the number of vertices, and density. We classify the graphs that do not fit the GPU memory as large-scale and display them separately in Table~\ref{table:graph_summary}.\looseness=-1

\begin{table}[h]
\caption{\small{Medium- and large-scale graphs used.}}
\label{table:graph_summary}
\centering
\scalebox{0.98}{
\begin{tabular}{lrrr}
\textbf{Graph}  & $|V|$ & $|E|$ & \textbf{Density} \\
\hline
{\tt com-dblp}~\cite{snapnets}        & 317,080      & 1,049,866      & 3.31  \\         
{\tt com-amazon}~\cite{snapnets}       & 334,863      & 925,872      & 2.76 \\     
{\tt youtube}~\cite{yt}         & 1,138,499    & 4,945,382    & 4.34             \\ 
{\tt flickr}~\cite{yt}     & 1,715,254  &  22,613,981  &     13.18 \\
{\tt com-orkut}~\cite{snapnets}        & 3,072,441    & 117,185,083   &  38.14 \\
{\tt com-lj}~\cite{snapnets}           & 3,997,962     &  34,681,189  &  8.67  \\
\hline
{\tt hyperlink2012}~\cite{hl}   & 39,497,204    & 623,056,313       & 15.77 \\
{\tt twitter\_{rv}}~\cite{nr}      & 41,652,230    &  1,468,365,182    & 35.25 \\
{\tt com-friendster}~\cite{snapnets}   & 65,608,366    &   1,806,067,135   &   27.53 \\
\end{tabular}
}
\end{table}

\noindent{\bf{\underline{Data preparation for ML tasks}}}: 
For link prediction, we split the input $G$ into two; $G_{train} = (V_{train}, E_{train})$ with $80\%$ of the edges, and $G_{test} = (V_{test}, E_{test})$ with the remaining $20\%$. In order to ensure $V_{test} \subseteq V_{train}$, we remove all the isolated vertices in $G_{train}$, and also all the edges $(u,v) \in G_{test}$ where $u \notin G_{train}$ or $v \notin G_{train}$. We generate the embedding matrices using $G_{train}$ employing the tools mentioned above. Then we utilize a Logistic Regression model to predict the existence of the edges in $G_{test}$. To train the model, we generate $|E_{train}|$ amount of positive and negative samples. The positive samples are directly taken from $E_{train}$, and the negative samples are randomly selected from $(V_{train} \times V_{train}) \setminus E_{train}$. A sample is generated by doing an element-wise multiplication of ${\mathbf M}_0[u]$ and ${\mathbf M}_0[v]$, where $(u,v)$ is selected as a training sample. The same procedure is used to generate the test samples. Finally, we report the \emph{AUCROC} score obtained from the test set~\cite{roc}. For medium-scale graphs, we use  {\tt LogisticRegression} from {\tt scikit-learn}. However, logistic regression becomes too expensive for large-scale graphs. Thus, for such graphs, we use {\tt SGDClassifier} from {\tt scikit-learn} with a logistic regression solver. \looseness=-1

For node classification, we train a {\em one-vs-rest} multi-label logistic regression classifier from {\tt scikit-learn}. We limit our evaluation scope to the top $100$ labels on all graphs and report Macro-, and Micro-F1 scores using $10\%$ labeled data. Since the label information is hidden from the embedding process, the entire $G$ is used for embedding. The embedding vector of a vertex $v$, ${\mathbf M}_0[v]$, and its corresponding labels are directly used as samples without any preprocessing.\looseness=-1

\noindent{\bf{\underline{State-of-the-art tools used for evaluation}}}:
The following state-of-the-art tools are used as the baselines for \malgo.

\begin{itemize}[leftmargin=*]
\item \versex is a CPU parallel graph embedding tool~\cite{verse18} that contains three different vertex-similarity measures, including adjacency similarity and personalized page rank (PPR). For the experiments, we use PPR with an $\alpha = 0.85$, and $0.0025$ as the learning rate. For link prediction, we run with $e = 600, 1000$, and $1400$ and report the best AUCROC. For node classification, we use $e = 1000$.

\item \graphvite~\cite{graphvite19} (GraphVite) is a multi-GPU embedding tool that matches the embedding quality of the CPU based tools while being much faster. We created two settings for the tool used on medium scaled graphs, 
{\em fast} and {\em slow}, with $e = 600$ and $1000$, respectively. For large-scale graphs, we use $e=100$. For the remaining hyper-parameters, we use the authors' recommended values and LINE as the embedding method. The tool can not embed graphs with {{$|V| > 12M$ on a single GPU~\cite{graphvite19}.}}

\item \mile is a graph embedding tool that applies a multi-level algorithm with coarsening. Unlike other multi-level embedding tools, \mile only does training at the coarsest level and projects the coarsest graph's embedding to that of the original graph using a neural network. We use {\sc DeepWalk} as the embedding method, {\sc MD-GCN} as the projection method, $8$ levels of coarsening, and a learning rate of $0.001$. \mile does not allow configuring the number of epochs.\looseness=-1

\item \pbg \cite{pbg19} (PyTorch-BigGraphs) is a popular, multi-relation embedding tool that can work with distributed environments via the \emph{torch.distributed} back-end. It recently started to support GPU-based embedding. 
We use the default hyper-parameters set by the authors except the number of epochs which is set as $e = 20$ for link prediction, $e = 30$ for node classification, and $e = 5$ for large-scale graphs.\looseness=-1
\end{itemize}

\begin{table}[!h]
\caption{\small{\malgo configurations, {\em fast}, {\em normal}, and {\em slow} for medium- and large-scale graphs. A version with no coarsening is also used in the experiments.}}
\label{table:configs}
\centering
\scalebox{0.98}{
\begin{tabular}{l|rr|r|r}
\textbf{Configuration} & $p$ & $lr$ & $e_{medium}$ & $e_{large}$\\
\hline
{\em fast}   & 0.1 & 0.050 & 600 & 100 \\
{\em normal} & 0.3 & 0.035 & 1000 & 200 \\
{\em slow}   & 0.5 & 0.025 & 1400 & 300 \\
\hline
{\em NoCoarse} & - & 0.045 & 1000 & 200  \\
\end{tabular}
}
\end{table}
We report four different configurations of \malgo; {\em NoCoarse}, {\em fast}, {\em normal}, and {\em slow}. We report the details of these configurations in Table~\ref{table:configs}. From {\em slow} to {\em fast}, we decrease the smoothing ratio and the number of epochs and increase the learning rate for compensation. We report the {\em NoCoarse} version, which does not apply coarsening to understand the impact of coarsening explicitly. Since the epoch definitions of the respective tools are not aligned, we define a single epoch as performing $|E|$ updates, given by \graphvite~\cite{graphvite19}, and adjust epochs accordingly for fairness.
\looseness=-1


\begin{table}[!h]
\centering
\caption{\small{Performance of \malgo coarsening without any heuristic~({\bf naive}), with {\bf ordering}, and both with  {\bf ordering} and {\bf hub$^2$-restriction}. {For each level $i$, the time to coarsen the graph $T_{crs}$, as well as the number of vertices $|V_i|$ of graph $i$, and its average degree. We also show the total training time $T_{train}$ after coarsening. {Results are obtained on Arch-2 with 16 threads.}}}}
\label{table:coarsening_optimizations}
\setlength{\tabcolsep}{4pt}
\scalebox{0.97}{
\begin{tabular}{c@{\hspace{1em}}c@{\hspace{1em}}r@{\hspace{1em}}rrrrrr}
  &  & \textbf{AUC} & & &  &  & \\
 \textbf{$G$} & \textbf{Heuristics} & \textbf{ROC} & \textbf{$T_{train}$ (s)} & \textbf{$i$} & \textbf{$T_{crs}$ (s)} & $|V_{i}|$ & $\frac{|E_{i}|}{|V_{i}|}$ \\
\hline
    \multirow{9}{*}{\rotatebox[origin=c]{90}{{\tt com-friendster}}} 
    & \multirow{2}{*}{{\bf naive}} & \multirow{2}{*}{0.964} & \multirow{2}{*}{43260.0} & 1 & 80.6 & 62,603K & 46.2 \\
    & & & & 2 & 16.0 & 28,690K & 64.2 \\ \cline{2-8}
    & \multirow{2}{*}{+{\bf ordering}} & \multirow{2}{*}{0.921} & \multirow{2}{*}{44912.4} & 1 & 62.0 & 62,603K & 46.2 \\
    & & & & 2 & 9.5 & 26,666K & 44.7 \\ \cline{2-8}
    & \multirow{5}{*}{+{\bf hub$^2$-rest.}} & \multirow{5}{*}{0.972} & \multirow{5}{*}{2316.5} & 1 & 133.1 & 62,603K & 46.2 \\
    & & & & 2 & 98.9 & 20,410K & 131.7 \\ 
    & & & & 4 & 7.36 & 1,298K & 842.5 \\ 
    & & & & 8 & 0.06 & 3.2K & 3019.5 \\
    & & & & 11 & - & 0.8K & 822 \\
\end{tabular}
}
\vspace*{-2ex}
\end{table}

\subsection{Experiments on coarsening performance} \label{coarsening-performance}

Recently, we showed that the coarsening approach used has a significant impact on the embedding performance, both in terms of accuracy and runtime. For completeness, here we also briefly analyze the coarsening phase. For a detailed evaluation, we refer the reader to~\cite{understanding}. We first use the coarsening algorithm {\bf naive}ly with no heuristics, then only with {\bf ordering}, and lastly with both {\bf ordering} and {\bf hub$^2$-restriction} and report the results in Table~\ref{table:coarsening_optimizations}. We then investigate the impact of coarsening depth on embedding and report the results in Table~\ref{table:coarsening_levels}.\looseness=-1

The {\bf naive} variant displays a poor performance regarding both coarsening efficiency and runtime. Although we observe a slight improvement in coarsening efficiency with {\bf ordering}, the embedding quality degrades substantially, i.e., from $0.964$ to $0.921$ for {\tt com-friendster}. We believe sorting amplifies an already existing problem; as explained in Section~\ref{s_coarsening}, the algorithm ends up with a couple of huge hub-vertices that prevent further coarsening. With {\bf hub$^2$-restriction}, this problem is mitigated; the largest graph in our dataset, {\tt com-friendster}, is reduced from $63$M vertices to $823$ vertices in only $10$ iterations. Furthermore, the heuristics positively affect the embedding quality; we observe a $1.5\%$ increase in AUCROC on average. These show that a carefully designed coarsening algorithm is critical for both speed and accuracy.\looseness=-1

\begin{table}[htbp]
\setlength{\tabcolsep}{5pt}
\centering
\caption{\small{The embedding performance with various coarsening depths $D \in \{3, 5, 7\}$ on Arch-2. The timings are given in seconds and the AUCROC scores are given as percentage.}}
\label{table:coarsening_levels}
\scalebox{0.96} {
\begin{tabular}{l|rr|rr|rr}
{\bf Graph} & ($\mathbf{D = 3}$) & AUC  & ($\mathbf{D = 5}$) & AUC & ($\mathbf{D = 7}$) & AUC\\
& Time & ROC & Time & ROC & Time & ROC\\
\hline
{\tt com-dblp} & 11.1 & 97.7 & 5.3 & 97.7 & 3.1 & 97.9 \\
{\tt com-amazon} & 10.5 & 98.0 & 5.1 & 98.4 & 3.0 & 98.5 \\
{\tt youtube} & 45.1 & 97.2 & 20.0 & 97.5 & 11.3 & 98.0 \\
{\tt com-lj} & 365.3 & 97.3 & 157.5 & 97.8  & 86.6 & 98.5 \\
{\tt com-orkut} & 1117.9 & 97.5  & 474.4 & 97.9  & 263.2 & 98.3 \\
\end{tabular}
}
\end{table}

As described in Section \ref{sec:gosh}, when $D$, the number of coarsening levels increases, the amount of work reserved for higher levels decreases. Naturally, with larger $D$, \malgo gets faster. One may expect that this embodies a trade-off between speed and accuracy. On the contrary, as shown in Table~\ref{table:coarsening_levels}, the AUCROC scores tend to improve. In fact, for all of the graphs, we observe a consistent increase (by an average of $0.7\%$) in AUCROC as $D$ increases from $D=3$ to $D=7$. This phenomenon shows that the updates on the coarser levels have a significant {positive} impact on embedding quality. {We believe, with the cumulative effect of the updates on the deeper levels, coarsening is able to guide the embedding algorithm to find better local minimas.} However, there is no strict relationship between AUCROCs and $D$. Indeed, with many levels, most of the epochs will be spent on similar graphs due to the exponential nature of \malgo's epoch distribution. We found that around $10$ levels, where each level is at least $80\%$ smaller than its predecessor, yield a good performance.\looseness=-1 

\begin{table}[!h]
\centering
\caption{\small{\mile vs \malgo coarsening on {\tt com-orkut}. $\tau = 16$ threads on Arch-2 are used for \malgo. }}
\label{table:MILEvsO}
\setlength{\tabcolsep}{4pt}
\scalebox{0.98}{
\begin{tabular}{crrr|crrr}
\textbf{Alg.} & $i$ & \textbf{Time (s)} & \textbf{$|V_{i}|$} & \textbf{Alg.} & $i$ & \textbf{Time (s)} & \textbf{$|V_{i}|$} \\
\hline
\multirow{5}{*}{\rotatebox[origin=c]{90}{\mile}} & 0 & - & 3056838 & \multirow{5}{*}{\rotatebox[origin=c]{90}{\malgo}} & 0 & - & 3056838 \\
             &2 & 237.39 & 768804 && 2 & 1.23 & 213707 \\
            &4 & 151.24 & 192507 && 4 & 0.16 & 8084 \\
            &6 & 128.47 & 48183 && 6 & 0.01 & 701 \\
            &8 & 99.73 & 12062 && 8 &  $<$ 0.01  & 275 \\\hline
            &Tot. & 1308.31 & - && Tot. & 6.60 & - \\
\end{tabular}
}
\vspace*{-2ex}
\end{table}

\noindent{\bf{\\\underline{\malgo vs \mile}:}}
As the last experiment on coarsening, we compare \mile and \malgo in Table~\ref{table:MILEvsO} on graph {\tt com-orkut} that has $3$ million vertices and $100$ million edges. \malgo's parallel coarsening uses $16$ threads, yet \mile's coarsening is not parallel. Since \mile does not have a stopping criterion for coarsening, we use $8$ levels of coarsening for both algorithms. \malgo is more efficient than \mile regarding the number of vertices obtained in each level while being $264$ times faster. For instance, \malgo shrinks the graph into 230 vertices, where \mile obtains 12,062 vertices. As our experiments show, a high-performance, parallel coarsening with a high efficiency and effectiveness is particularly important.

\begin{table*}[!t]
\setlength{\tabcolsep}{3pt}
\centering
\caption{\small{Link prediction results on medium-scale graphs. \versex and \malgo uses $\tau = 16$ threads on Arch-1. \mile is a sequential tool. GPU-based tools use the V100 GPU. The speedup values are given w.r.t. \versex. AUCROC scores are in percentage.}}
\label{tab:medium_results}
\scalebox{0.95}{
\begin{tabular}{l||c|rrr||c|rrr||c|rrr||c|rrr}
      & &  &   & {\bf AUC} & &  &  & {\bf AUC} & & & & {\bf AUC} & & & & {\bf AUC}   \\
     {\bf Algorithm} & $G$ & {\bf Time (s)} & {\bf Speedup} & {\bf ROC} & $G$ & {\bf Time (s)} & {\bf Speedup} & {\bf ROC} & $G$ & {\bf Time (s)} & {\bf Speedup} & {\bf ROC} & $G$ & {\bf Time (s)} & {\bf Speedup} & {\bf ROC} \\
     \hline\hline
     \versex & 
     \multirow{8}{1em}{\rotatebox[origin=c]{90}{\tt com-dblp}} & 128.02 & 1.0$\times$ &  97.76 & 
     \multirow{8}{1em}{\rotatebox[origin=c]{90}{\tt youtube}} & 700.26 & 1.0$\times$ & \textbf{97.99} & 
     \multirow{8}{1em}{\rotatebox[origin=c]{90}{\tt com-lj}} &7472.89 & 1.0$\times$ & 98.91  & 
     \multirow{8}{1em}{\rotatebox[origin=c]{90}{\tt com-orkut}} & 25020.28 &1.0$\times$  &  \textbf{98.28}\\
     \cline{1-1}\cline{3-5}\cline{7-9}\cline{11-13}\cline{15-17}
     \mile & & 122.55 & 1.04$\times$ & 97.70 &
           &1847.18  & 0.38$\times$ & 94.55 & & 3046.88 & 2.45$\times$ & 85.89 & & 9347.67$^{\mathrm{a}}$ & 2.68$\times$ & 90.22 \\ 
     \cline{1-1}\cline{3-5}\cline{7-9}\cline{11-13}\cline{15-17}
     \graphvite-fast & & 5.96 & 21.47$\times$ & 95.72 & & 22.48 & 31.15$\times$ & 97.12 & & 146.21 & 51.11$\times$ & 98.33 & & 456.00 & 45.87$\times$ & 97.75\\ 
     \graphvite-slow & & 8.80 & 14.56$\times$ & 97.48 & & 34.96 & 20.19$\times$ & 97.08 & & 236.00 & 31.66$\times$ & 98.36 & & 746.25 & 33.53$\times$ &  97.30 \\ 
     \cline{1-1}\cline{3-5}\cline{7-9}\cline{11-13}\cline{15-17}
     \pbg & & 21.29 & 6.01 $\times$ & 97.73 & & 74.42 & 9.41$\times$ & 97.07  & & 497.06 & 15.03 $\times$ & 98.40 & & 1102.90 & 22.69$\times$ & 98.42 \\ 
     \cline{1-1}\cline{3-5}\cline{7-9}\cline{11-13}\cline{15-17}
     \malgo-fast & & 0.53  & 241.44$\times$& 97.02 & & 1.89 & 370.88$\times$ & 97.67 & & 11.38 & 656.55$\times$ & 98.28 & & 30.52 & 819.68$\times$ & {\bf 98.86}\\
     \malgo-normal & & 1.41  & 90.95$\times$& {\bf 97.79} & & 6.11 & 114.65$\times$ & {\bf 98.00} & & 39.44 & 189.48$\times$ & {\bf 98.74} & & 211.64 & 211.64$\times$ & 98.66\\
     \malgo-slow & & 2.83  & 45.20$\times$& {\bf 97.95} & & 12.64 & 55.39$\times$ & {\bf 98.00} & & 85.26 & 87.65$\times$ &  98.72 & & 261.57 & 95.66$\times$ & 98.37\\
     \malgo-NoCoarse & & 19.17 & 6.68$\times$ & 97.13 & & 90.46 & 7.74$\times$ & 96.51 & & 655.00 & 11.41$\times$ & 96.57 & & 2228.92 & 11.23$\times$ & 96.88 \\ 
     \hline
     \multicolumn{15}{l}{$^{\mathrm{a}}$We get these results by embedding with 8 threads instead of 16 due to an OOM issue with 16 threads.}
\end{tabular}
}
\end{table*}

\begin{table*}[htbp]
    \centering
     \caption{\small{Node classification results on all labeled graphs. Each value is an average of 3 embeddings and 5 evaluations. Timings are given in sec. and Micro-, Macro-F1 as percentages. {Experiments are performed on Arch-1 using a V100 GPU.}}}
    \label{tab:node_class_all}
    \scalebox{0.95}{
        \begin{tabular}{l || c | rrr || c | rrr || c | rrr}
            \multirow{2}{*}{\textbf{Algorithm}} & $G$ & \multirow{2}{*}{\textbf{Time (s)}} & \textbf{Micro} & \textbf{Macro} & \multirow{2}{*}{$G$} & \multirow{2}{*}{\textbf{Time (s)}} & \textbf{Micro} & \textbf{Macro1} & \multirow{2}{*}{$G$} & \multirow{2}{*}{\textbf{Time (s)}} & \textbf{Micro} & \textbf{Macro1} \\
            & & & \textbf{F1} & \textbf{F1} & &  & \textbf{F1} & \textbf{F1} & &  & \textbf{F1} & \textbf{F1} \\
            \hline\hline
            \versex & \multirow{7}{1em}{\rotatebox[origin=c]{90}{{\tt com-dblp}}} & 166.21 & 46.97 & 43.27 & \multirow{7}{1em}{\rotatebox[origin=c]{90}{{\tt com-amazon}}}& 146.34 & 94.94 & 94.63 & \multirow{6}{1em}{\rotatebox[origin=c]{90}{{\tt com-lj}}}& 4893.53 & 81.99 & 81.24 \\
            \cline{1-1}\cline{3-5}\cline{7-9}\cline{11-13}
            \graphvite-fast & & 7.13 & 53.95 & 33.83 &  & 6.39 & 72.97 & 68.91 & & 180.71 & 87.25 & 84.64 \\
            \graphvite-slow & & 10.56 & 57.05 & 42.90 & & 9.43 & 90.16 & 89.03 & & 290.49 & \textbf{87.68} & \textbf{85.59} \\
            \cline{1-1}\cline{3-5}\cline{7-9}\cline{11-13}
            \pbg & & 32.79 & 39.38 & 35.19 & & 29.63 & 89.02 & 88.51 & & 935.18 & \textbf{87.94} & \textbf{86.54} \\
            \cline{1-1}\cline{3-5}\cline{7-9}\cline{11-13}
            \malgow-fast & & 0.86 & 52.82 & 49.38 & & 0.77 & 97.22 & 96.43 & & 17.68 & 84.68 & 83.34 \\
            \malgow-normal & & 1.76 & 60.36 & 58.76 & & 1.64 & \textbf{98.21} & \textbf{97.80} & & 60.08 & 85.60 & 84.26 \\
            \malgow-slow & & 3.41 & \textbf{62.27} & \textbf{61.22} & & 2.69 & 98.31 & 97.82 & & 127.53 & 86.36 & 85.04 \\
            \malgow-NoCoarse & & 15.62 & 58.59 & 54.10 & & 13.86 & 97.44& 97.34 & & 541.43 & 86.03 & 84.63 \\
            \cline{1-1}\cline{3-5}\cline{7-9}\cline{11-13}
            \hline\hline
            \versex & \multirow{6}{1em}{\rotatebox[origin=c]{90}{{\tt youtube}}} & 882.29 & 24.70 & 16.23 & \multirow{6}{1em}{\rotatebox[origin=c]{90}{{\tt flickr}}} & 4135.20 & 40.15 & 37.27 & \multirow{6}{1em}{\rotatebox[origin=c]{90}{{\tt com-orkut}}} & 24917.93 & 68.03 & 65.24 \\
            \cline{1-1}\cline{3-5}\cline{7-9}\cline{11-13}
            \graphvite-fast &  & 26.24 & 34.51 & 24.90 &   & 110.29 & 48.77 & 46.62 &   & 564.42 & 83.98 & 83.26 \\
            \graphvite-slow & & 41.04 & 35.27 & 25.77 & & 177.58 & 48.54 & 46.05 & & 926.09 & \textbf{84.38} & \textbf{83.74} \\
            \cline{1-1}\cline{3-5}\cline{7-9}\cline{11-13}
            \pbg & & 134.11 & 34.29 & 23.54 & & 597.72 & 47.20 & 44.68 & & 3111.03 & \textbf{82.79} & \textbf{81.96} \\
            \cline{1-1}\cline{3-5}\cline{7-9}\cline{11-13}
            \malgow-fast & & 2.62 & 32.33 & 23.13 & & 8.74 & 46.43 & 44.63 & & 39.49 & 77.23 & 75.91 \\
            \malgow-normal & & 6.86 & 34.70 & 25.30 & & 30.12 & 49.57 & 47.88 & & 155.55 & 81.63 & 80.75 \\
            \malgow-slow & & 13.63 & \textbf{35.30} & \textbf{25.79} & & 65.64 & 50.24 & 48.54 & & 342.62 & 82.78 & 82.02 \\
            \malgow-NoCoarse & & 73.54 & 35.07 & 25.33 & & 342.54 & \textbf{50.77} & \textbf{49.10} & & 1836.23 & 83.20 & 82.39 \\
        \end{tabular}
    }
\end{table*}

\subsection{{Experiments on epoch synchronization}}\label{sec:ep_sync}

\begin{figure}[!h]
    \centering
    \includegraphics[width=0.97\linewidth]{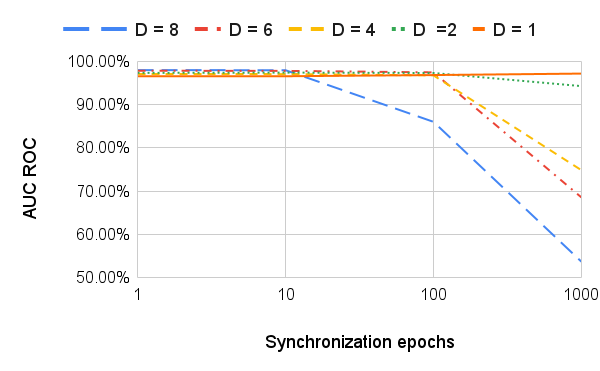}
    \caption{\small{AUCROC scores on {\tt youtube} using different epoch synchronization periods and the V100 GPU of Arch-1. Each line  represents a different coarsening depth, with $D = 1$ representing an embedding without any coarsening.}}
    \label{fig:sync}
    \vspace*{-2ex}
\end{figure}

{As mentioned in Section~\ref{sec:gpu_imp}, \malgo carries out the updates of a single training epoch completely in parallel. Yet, it serializes the training epochs to avoid further race conditions. Figure~\ref{fig:sync} shows the effect of increasing epoch parallelism on the AUCROC score of link prediction on {\tt youtube}. Each point on the $x$-axis represents the number of epochs executed in parallel on the GPU without synchronization, with $1$ being the current implementation with sequential epochs, and $1000$ being the other extreme, where all training epochs are executed with no synchronization. 

Fig.~\ref{fig:sync} shows different coarsening granularities with different lines, where $D = 1$ implies no coarsening, and $D = 8$ is the maximum number of coarsening levels. Although epoch synchronization has little to no effect with a small number of coarsening steps, its benefit heaves in sight as the number of levels increases and graphs become smaller in lower levels; a small number of vertices implies higher probability for potential race conditions. These results clearly demonstrate the importance of synchronizing epochs - especially in a multi-level setting. They also show that multi-level, GPU-based embedding tools may be improved in terms of runtime for large graphs by allowing a number of concurrent epochs on the upper levels but forcing full synchronization elsewhere. 
We leave this as a future study.\looseness=-1}

\subsection{Experiments on medium-scale graphs}

Tables~\ref{tab:medium_results},~\ref{tab:node_class_all}, and~\ref{tab:node_class_youtube} provide the runtimes, AUCROC~(for link prediction), and Micro-, and Macro-F1 scores~(for node classification), for the tools evaluated on medium-scale graphs. In addition to \malgo~(for link prediction) and \malgow~(for node classification), we use \graphvite, \mile, \pbg, and \versex.\looseness=-1 

\begin{table}[htbp]
    \caption{\small{Node classification results on {\tt youtube} w.r.t. labeled node percentages on Arch-2. Each value is the average of 3 embeddings and 5 evaluations. Both \graphvite and \malgow use the {\em slow} configuration.}}
    \centering
        \begin{tabular}{c|c|ccccc}
            {\bf F1 \%} & {\bf Algorithm} & {\bf 2\%} & {\bf 4\%} & {\bf 6\%} & {\bf 8\%} & {\bf 10\%} \\
            \hline
            \multirow{3}{1em}{\rotatebox[origin=c]{90}{\bf Micro}}& \versex & 22.05 & 22.15 & 22.92 & 22.81 & 23.00\\
            & \graphvite & \textbf{33.23} & 34.06 & \textbf{34.83} & 35.50 & 35.90\\
            & \malgow & 32.37 & \textbf{34.08} & 34.73 & \textbf{35.52} & \textbf{36.14}\\
            \hline
            \multirow{3}{1em}{\rotatebox[origin=c]{90}{\bf Macro}} & \versex & 12.60 & 13.85 & 14.77 & 14.66 & 14.93\\
            & \graphvite & 20.23 & 23.24 & \textbf{24.82} & \textbf{25.48} & 26.04\\
            & \malgow & \textbf{20.32} & \textbf{23.60} & 24.45 & 25.03 & \textbf{26.21}
        \end{tabular}
    \label{tab:node_class_youtube}
    \vspace*{-2ex}
\end{table}

To evaluate the runtime performance on link prediction, we use \versex as the baseline and present the speedups of each tool in Table~\ref{tab:medium_results}. The results show that \malgo-fast is an ultra-fast solution that produces very accurate embeddings at a fraction of the time compared to all the systems under evaluation. It can achieve speedups over \versex of nearly three orders of magnitude while sacrificing only an average of $0.28\%$ in AUCROC. When compared to \mile, it is superior in terms of AUCROC in all four graphs while being at least two orders of magnitude faster. Although \pbg is on par with \malgo-fast on link prediction quality, it is $40 \times$ slower on average. Finally, when compared to \graphvite, we find that \malgo-normal is two orders of magnitude faster and scores $0.06\%$ higher AUCROC on average, setting the new state-of-the-art in link prediction.\looseness=-1

Relative to link prediction, node classification is a more challenging problem. For both \malgow and \graphvite, node classification requires more work to converge, where {\em fast} configurations are not as successful as they are for link prediction. The adjacency similarity measure, which respects only the first-order proximity, is sufficient for link prediction; however, a more sophisticated measure is a must for classification. For this purpose, we experimented with \versex using the personalized page rank (PPR) similarity measure, and report the results in Tables~\ref{tab:node_class_all} and~\ref{tab:node_class_youtube}. Overall, \versex performs poorly with respect to \graphvite and \malgow, which shows that PPR similarity measure also falls short for node classification. We found that the best option for node classification is random walks~\cite{deepwalk}. The biggest difference between random walks and PPR is related to the number of samples collected from a single walk. Compared to a single sample per walk by PPR, with~\cite{deepwalk}, tens of samples are collected from a single walk depending on the walk distance and sampling window size. These results highlight the importance of coherent sampling.\looseness=-1


We compare \malgow with the state-of-the-art, \graphvite and \pbg, using the best Micro- and Macro-F1 scores for both tools. As Table~\ref{tab:node_class_all} shows,  \malgow-{\em slow} achieves an average improvement of $1.88\%$-$4.38\%$ and $5.63\%$-$6.59\%$ in Micro- and Macro-F1 scores while achieving an average speed-up of $2.7\times$ and $9.3\times$ with respect to \graphvite and \pbg. Although \malgow-{\em fast} obtains significant speedups over \graphvite and \pbg, it falls short on node classification accuracy on all the graphs besides {\tt com-amazon}. Furthermore, \malgo could saturate with the {\em normal} configuration on link prediction; however \malgow, on node classification, requires more training to converge.  For instance, switching from {\em fast} to {\em normal} results in an average AUCROC score increase of $0.3\%$ in link prediction~(Table~\ref{tab:medium_results}), and Micro and Macro-F1 increase of $3.2\%$ and $3.0\%$ in node classification, while reducing the speed on average only by a factor of $3\times$~(Tables~\ref{tab:medium_results} and~\ref{tab:node_class_all}).\looseness=-1

{Table~\ref{tab:node_class_youtube} shows the node-classification results with different percentages of labeled nodes. In terms of Micro- and Macro-F1 scores, unlike link prediction, \graphvite and \malgow are in the same ballpark. Hence, coarsening does not have the same positive impact as on link prediction on node classification. However,  even for node classification, coarsening is still needed for a fast embedding.\looseness=-1}

   


\subsection{Experiments on handling large graphs}\label{lges}

\begin{figure}[!h]
    \centering
    \includegraphics[width=0.95\linewidth]{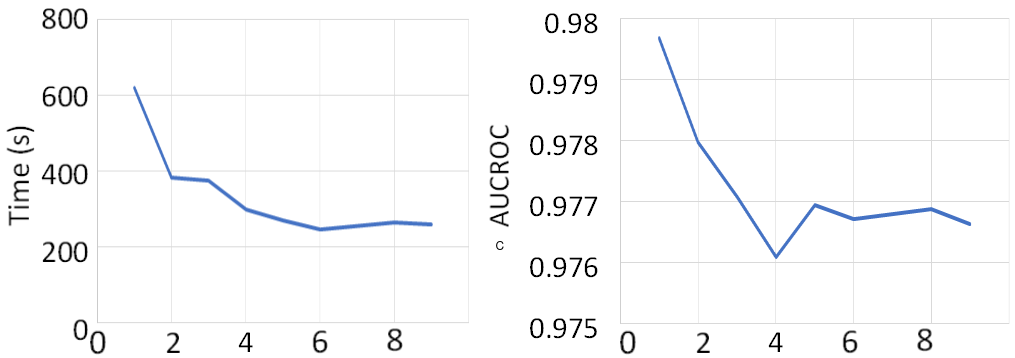}
    \caption{\small{The runtimes~(left) and AUCROC scores~(right) for large-scale graph embedding on {\tt hyperlink} with different $B$ values. {The results are acquired using a single Titan X GPU on Arch-2.}}}
    \label{fig:bvauctime}
    \vspace*{-2ex}
\end{figure}

Figure~\ref{fig:bvauctime} demonstrates the effect of adjusting the number of positive samples per vertex in a single sample pool, i.e., $B$, for large-graph embedding. The figures show the runtimes in seconds~(left) and the  link prediction AUCROC scores~(right). Increasing $B$ leads to a trade-off between embedding performance and quality; an increase in $B$ decreases the runtime due to the reduction in number of embedding rounds and hence, number of submatrix copies. However, increasing $B$ also results in more updates to be carried out on subsets of the graph {\em in isolation} from the remaining subsets. Thus, it delays the propagation of these updates and hurts the accuracy. In \malgo, we use a default value of $B = 5$ as it achieves a performance boost without a significant negative impact on the accuracy.

\begin{figure}[h!]
    \centering
    \includegraphics[width=0.85\linewidth]{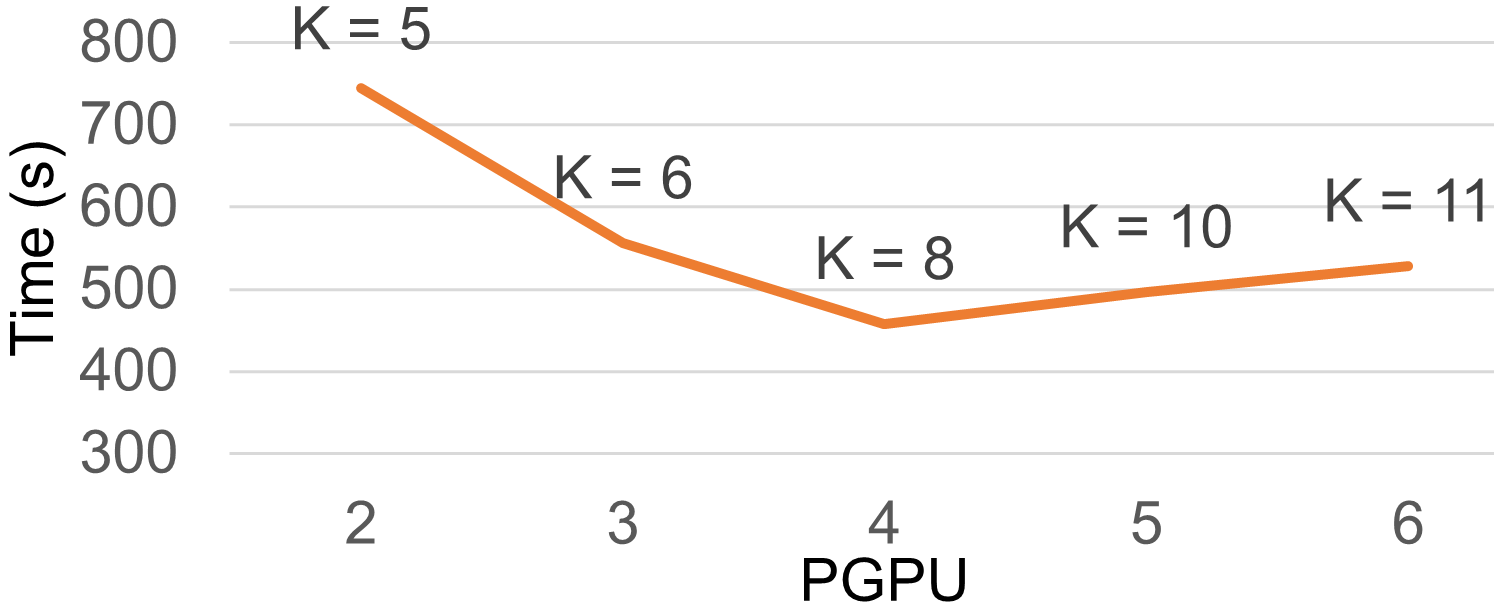}
    \caption{\small{Time taken for 100 epochs on {\tt hyperlink2012} with different $P_{GPU}$ (and $K$) values. Experiments are performed with $S_{GPU}=4$ and $B=5$ on a single Titan X on Arch-2.}}
    \label{fig:time_pgpu}
    \vspace*{-1ex}
\end{figure}

{Figure~\ref{fig:time_pgpu} shows the relationship between the number of submatrices kept on the GPU, i.e., $P_{GPU}$, the number of parts $V$ is partitioned into, i.e., $K$, and the embedding time. When $P_{GPU}$ increases from $2$ to $4$, the embedding time is reduced drastically. This is because the embedding kernels are overlapped with the submatrix transfers between the host and the GPU. However, the reduction stops after $P_{GPU} = 4$. We suspect that this is due to the increase in $K$. To elaborate, as the number of submatrices on the GPU increases, the size of a single submatrix decreases and hence, the number of parts increases. More parts translate to more submatrix communications per embedding round, and the increased communication cost deteriorates the overall performance.}
\looseness=-1

\begin{table}[htbp]
\setlength{\tabcolsep}{3pt}
  \centering
  \caption{\small{Link prediction (LP) and node classification (NC) results on large-scale graphs. Every data-point is the average of 3 embeddings. 
  {Single- and multi-GPU experiments are conducted on Arch-1 and Arc-3, respectively.}}} 
  \label{tab:large_results-link}
  \scalebox{0.94}{
\begin{tabular}{crrrr}
      &  &  & & {\bf AUC }  \\
     {\bf Graph} & {\bf Algorithm} & {\bf LP Time (s)} && {\bf ROC (\%)}  \\
     \hline
     & \malgo-fast~(V100) & 78.00 & \multicolumn{2}{r}{97.37}  \\
     \small{{\tt hyperlink}} & \malgo-normal~(V100) & 247.19 & \multicolumn{2}{r}{99.21}  \\
     \small{{\tt 2012}} & \malgo-slow~(V100) & 564.57 & \multicolumn{2}{r}{\textbf{99.37}}  \\
     & \pbg~(V100) & 3762.49 & \multicolumn{2}{r}{97.27} \\
     & \graphvite~(3$\times$ A100) & 914.886 & \multicolumn{2}{r}{98.81} \\
     \hline
     & \malgo-fast~(V100)  & 81.65 & \multicolumn{2}{r}{95.35}  \\
      \small{{\tt twitter}} & \malgo-normal~(V100)  & 280.12 & \multicolumn{2}{r}{98.67}  \\
      \small{{\tt \_rv}} & \malgo-slow~(V100) & 681.78 & \multicolumn{2}{r}{\textbf{98.95}}  \\
     & \pbg~(V100) & 7072.71 & \multicolumn{2}{r}{91.88} \\
     & \graphvite~(3$\times$ A100)& 1726.79 & \multicolumn{2}{r}{98.67} \\
     \hline
     & \malgo-fast~(V100) & 171.03 & \multicolumn{2}{r}{88.10}  \\ 
    \small{{\tt com-}} & \malgo-normal~(V100) & 609.63 & \multicolumn{2}{r}{97.96}  \\
    \small{{\tt friendster}}  & \malgo-slow~(V100) & 1483.70 & \multicolumn{2}{r}{\textbf{98.73}} \\
     & \pbg~(V100) & 11770.59 & \multicolumn{2}{r}{98.12} \\
     & \graphvite~(3$\times$ A100)& 1660.12 & \multicolumn{2}{r}{98.01} \\
     \hline
       &  &  {\bf NC} & \textbf{Micro}& \textbf{Macro}  \\
     {\bf Graph} & {\bf Algorithm} & {\bf Time (s)} & \textbf{F1 (\%)}& \textbf{F1 (\%)}  \\
      \hline
      & \malgow-fast~(V100) &  371.54 & 61.63 & 58.45 \\ 
      \small{{\tt com-}}  & \malgow-normal~(V100) & 1665.69 & 70.92 & 68.96 \\
      \small{{\tt friendster}} & \malgow-slow~(V100)  & 4025.95 & 74.74 & \textbf{73.36} \\
     & \graphvite~(3$\times$ A100)& 3286.82 & \textbf{77.75} & 73.19\\
\end{tabular}
}
\end{table}

We report the results for large-scale graphs in Table~\ref{tab:large_results-link}.  
\versex timed out (12 hours) on all the large-scale graphs. \mile timed out on {\tt hyperlink2012} and could not embed the remaining two due to insufficient memory. Similar to \mile, \graphvite ran out of GPU memory~(on a single A100 GPU) on all the large-scale graphs. For this reason, we run it on the Arch-3 server equipped with four A100 GPUs. However, we used three out of four GPUs since \graphvite had (node-level) out-of-memory issues with four GPUs. As Table~\ref{tab:large_results-link} shows, although it is using a single V100, \malgo surpassed \graphvite's AUCROC scores~(on 3$\times$ A100s) in link prediction while being, on average, $3\times$ faster. Similarly, \malgo-slow is always better than \pbg in terms of AUCROC score and runtime.\looseness=-1 

For node classification, \malgow-slow on a single V100 is approximately $20\%$ slower compared to \graphvite using three A100s. In fact, \malgow-slow generates embeddings with similar F1 scores on a single A100 GPU approximately in 2000 seconds leveraging a larger, 80GB global memory. In terms of F1 scores, on {\tt com-friendster}, \malgow-slow obtains a worse Micro F1 score compared to \graphvite but have a slightly better Macro F1 score on average.\looseness=-1

\subsection{Experiments with smaller dimensions}

As explained in Section~\ref{sec:gpu_imp}, to utilize the GPU's full power, we use a warp-based implementation; a single warp performs a single update at a time. Since a warp contains $32$ threads; when $d < 32$, $32-d$ threads remain idle. In other words, \malgo takes almost same amount of time for all runs where $d < 32$. To increase efficiency, we apply the technique described in Section~\ref{sec:small_dimensions} for $d < 32$. With this assignment, as shown in Table~\ref{table:small_dimensions}, we observe $2.63\times$ and $1.8\times$ speedup for $d = 8, 16$ on {\tt com-orkut} compared to $d = 32$.\looseness=-1

\begin{table}[h!]
\centering
\caption{\small{\malgo's runtime with~(SM=+) \& without~(SM=-) small-dimension embedding and $\tau = 16$ threads. {Results are acquired using a single Titan X.}}}
\label{table:small_dimensions}
\scalebox{0.98}{
\begin{tabular}{ccrr|ccrr}
 \textbf{Graph} & \textbf{SM}& \textbf{$d$} & \textbf{Time (s)} & \textbf{Graph} & \textbf{SM}& \textbf{$d$} & \textbf{Time (s)}\\
\hline
    \multirow{6}{*}{\rotatebox[origin=c]{90}{{\tt com-orkut}}}  & \multirow{3}{*}{-} & 8 & 63.72 & \multirow{6}{*}{\rotatebox[origin=c]{90}{{\small{\tt soc-LiveJ.}}}}  & \multirow{3}{*}{-} & 8 & 40.13 \\
    & & 16 & 64.20 & & & 16 & 40.46 \\
    & & 32 & 64.95 & & & 32 & 41.22 \\\cline{2-4}\cline{6-8}
    & \multirow{3}{*}{+} & 8 & 24.27 & & \multirow{3}{*}{+} & 8 & 14.86\\
    & & 16 & 34.98 & & & 16 & 21.82 \\
    & & 32 & 64.54 & & & 32 & 40.93 \\
\end{tabular}
}
\vspace*{-3ex}
\end{table}

\subsection{Speedup breakdown}

\begin{figure}[h!]
    \centering
    \includegraphics[width=0.97\linewidth]{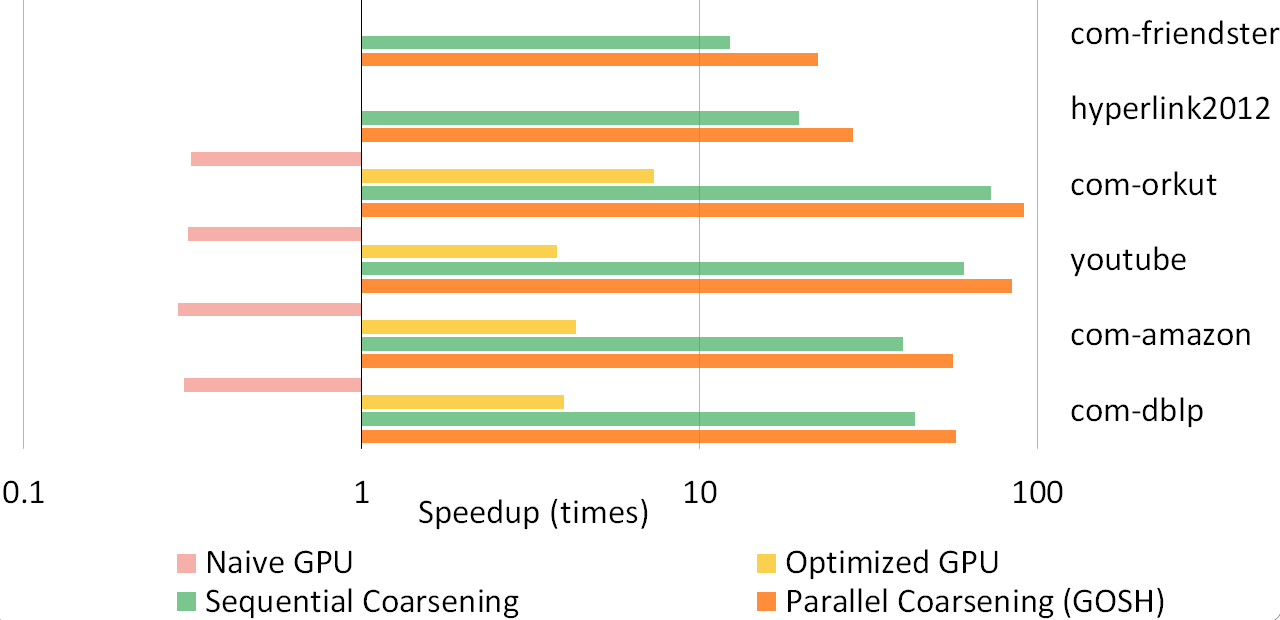}
    \caption{\small{Speedups of \malgo versions over a multi-core CPU implementation using 16 threads on Arch-2.}}
    \label{fig:speed}
    \vspace*{-2ex}
\end{figure}

{We run intermediate GPU-accelerated versions of \malgo and report their speedup over our optimized, 16-thread CPU implementation. We show the results in Figure~\ref{fig:speed}. In the figure, the first two results are for large-scale graphs, and the remaining four are for medium-scale graphs. We omit the results of the non-coarsened versions of large-scale graphs as they take a very long time. }

{The first version, {\em Naive GPU}, is 3.3$\times$ slower than the multi-core CPU version. The next, {\em Optimized GPU}, improves the first by leveraging GPU-based optimizations, e.g., memory coalescing, and utilizing shared memory to reduce global memory usage. It is, on average, 5.4$\times$ faster than the multi-core version. These versions do not use coarsening.}

{{\em Sequential Coarsening} introduces single-thread coarsening. 
It achieves an average speedup of 45$\times$ over the multi-core version and maintains the quality as shown in Tables~\ref{tab:medium_results}--\ref{tab:large_results-link}. 
Note that a single update on a super vertex in a coarsened graph is propagated to all the vertices it contains.} 

{The final version, {\em Parallel Coarsening}, employs multi-threaded coarsening. The improvement is more pronounced for larger graphs. For instance, sequential and parallel coarsening take $2469$ and $235$ seconds, respectively, on {\tt com-friendster}. Meanwhile, the total runtime of \malgo-normal is $2721$ secs. This means that parallel coarsening improves performance by $80\%$. However, it improves the runtime of {\tt com-dblp} by only $3\%$ since the embedding takes $2.2$ seconds, and the sequential and parallel coarsening times are $0.13$ and $0.06$ seconds, respectively.}

\section{Related Work}
\label{sec:rel}
In recent years, there has been a surge of interest in graph embedding. 
The literature is carefully documented in~\cite{Goyal2018GraphET} and can be categorized into three branches: factorization-, deep-learning-, and sampling-based embedding. The first class of tools factorizes {{the adjacency/}similarity matrices}~\cite{roweis2000nonlinear, lem,hope,grarep, prone, netsmf, homog}. Deep-learning-based approaches utilize deep autoencoders to reduce the dimensionality of the graph~\cite{sdne, dngr}. Lastly, in sampling-based embedding, a single-layer neural network is trained by choosing samples from the graph~\cite{verse18, LINE, deepwalk, node2vec}. Samples are chosen with respect to a similarity measure or a sampling strategy, varying throughout the works. Although there have been attempts to generalize the sampling process~\cite{verse18,hope}, a common ground has not been reached yet.

{Graph coarsening is often used to reduce the cost of large-scale graph problems. Most recently, Gilbert et al. presented CPU and GPU optimizations on many graph coarsening algorithms and demonstrated significant performance improvements on graph partitioning~\cite{coarse}. There have been attempts to use coarsening for embedding~\cite{HARP, mile18}, however, they do not utilize specialized processing units like GPUs. Distributed embedding approaches are also proposed to make the embedding  faster~\cite{pbg19, swivel, ordentlich2016network}.  Most notably, \cite{pbg19} can embed multi-relational graphs and utilizes a parameter partitioning schema similar to \malgo that provides distributed training on many multi-core and multi-GPU nodes. However, under hardware restrictions, such methods cannot be used for large-scale graph embedding.}

{Several studies have focused on accelerating graph algorithms by exploiting GPUs~\cite{graphvite19, edges, dgl_ke}. 
Both~\cite{graphvite19} and~\cite{edges} require the embedding matrix to fit in device memory, which constitutes a restriction on the graphs' size to be embedded when working with limited hardware.}
\malgo relaxes this limitation by leveraging a novel coarsening algorithm and a judiciously devised CPU and GPU parallel scheduling algorithm. Similar to \malgo, \cite{dgl_ke} is able to tackle limited-hardware restrictions by utilizing a memory partitioning schema that can embed large-scale graphs on a single GPU. However, we did not include it as a baseline since its knowledge-graph focused scoring functions make it incomparable to the other baselines in this work.\looseness=-1

\section{Conclusion and Future Work}
\label{sec:con}

This paper introduces a high-quality, fast, and flexible graph embedding algorithm \malgo{} that outperforms the state-of-the-art in terms of both efficiency and accuracy by exploiting a parallel, multi-level coarsening. 
\malgo can handle real-world, large-scale graphs with a single GPU via a partitioning schema and CPU sampling which are designed to minimize GPU idling. We achieve a new state-of-the-art on link prediction in terms of both speed and AUCROC. Moreover, we match the state-of-the-art on node classification in terms of F1-scores while being twice as fast.
We leave multi-node and multi-GPU support of \malgo as future work.

\ifCLASSOPTIONcaptionsoff
  \newpage
\fi

\section*{Acknowledgements}

This work was supported by Scientific and Technological Research Council of Turkey (TÜBİTAK) and EuroHPC Joint Undertaking through grant agreement No 220N254 and grant agreement No 956213 (SparCity), respectively. The numerical calculations reported in this paper were partially performed at TUBITAK ULAKBIM, High Performance and Grid Computing Center (TRUBA resources).



\renewcommand{\baselinestretch}{0.97}
\bibliographystyle{IEEEtran}
\bibliography{bare_adv}
\renewcommand{\baselinestretch}{1}

\vspace*{-2\baselineskip}
\begin{IEEEbiography}[{\includegraphics[width=1in,height=1.25in,clip,keepaspectratio]{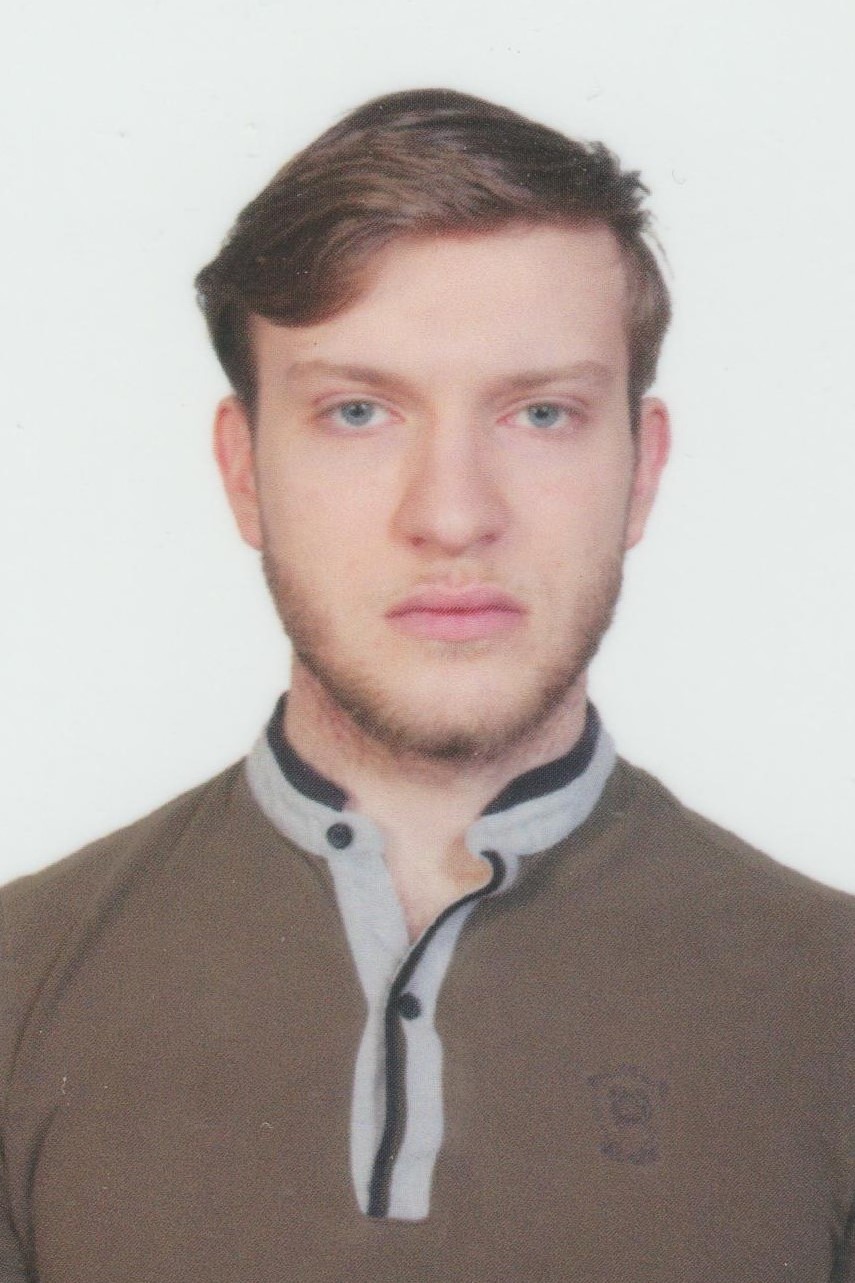}}]{Amro Alabsi Aljundi} is a Ph.D. student majoring in computer science and engineering at Sabanc{\i} University. He got his M.Sc. in Computer Science and Engineering at Sabancı University, and his bachelors in Computer Engineering at the American University of Sharjah. His research revolves around high performance computing and its applications on machine learning algorithms. 
    \end{IEEEbiography}
 \vspace*{-2\baselineskip}
\begin{IEEEbiography}[{\includegraphics[width=1in,height=1.25in,clip,keepaspectratio]{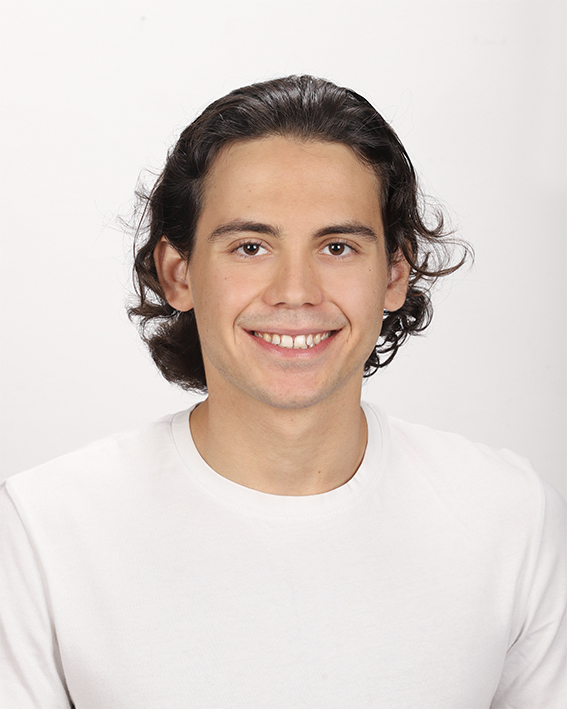}}]{Taha Atahan Aky{\i}ld{\i}z } is a Ph.D. student majoring in computer science and enginering at Sabanc{\i} University. He received both his B.S. and M.S. in Computer Science and Engineering from Sabancı University. His research interests include parallel programming, graph algorithms, and machine learning on graphs.
    \end{IEEEbiography}
 \vspace*{-2\baselineskip}
\begin{IEEEbiography}[{\includegraphics[width=1in,height=1.25in,clip,keepaspectratio]{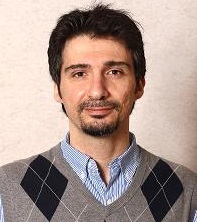}}]{Kamer Kaya} is an Asst. Prof. at Sabancı University. He got his PhD from Dept. Computer Science and Eng. from Bilkent University. He worked at CERFACS, France, as a post-graduate researcher. He then joined the Ohio State University in 2011 as a postdoctoral researcher, and in 2013, he became a Research Asst. Prof. in the Dept. of Biomedical Informatics.
His research interests include Parallel Programming and High Performance Computing.
    \end{IEEEbiography}





\end{document}